# Mapping Artistic Engagement with Quantum Technologies

Adrian Schmidt[1*], Oya Ok[2], Deniz İpek Zeynioğlu[3], Zeki C. Seskir[4]

[1]*Institute for Technology Assessment and Systems Analysis, Karlsruhe Institute of Technology, Berlin, Germany*
[2]*Department of Political Science and Public Administration, Bilkent University, Ankara, Turkey*
[3]*Uskudar American Academy, Istanbul, Turkey*
[4]*Institute for Technology Assessment and Systems Analysis, Karlsruhe Institute of Technology, Karlsruhe, Germany*

## Abstract

Quantum technologies (QT) are increasingly being taken up in artistic practices, for example in the form of visual artworks or quantum computer music. However, more comprehensive empirical studies on the temporal development and the different forms of QT-related art are still lacking. Therefore, this study conducted a systematic mapping and classification of QT-related artworks. To this end, a medium-independent corpus was established, and the artworks were classified into four categories: QT-Inspired, QT-Sourced, Simulated QT, and QT-Generated. This classification was LLM-assisted and checked against human-coded reference data as well as through manual review of unclear cases. In total, 281 QT-related artworks from 2010 to August 2025 were identified. The corpus shows a substantial increase in QT-related artworks since 2019, with a considerable share attributed to only a few artists. The different forms of QT relations coexist, while an increasing use of active hardware-based and simulated processes in artistic practices can be documented, particularly since 2021. Although no clear development trajectories between these categories can be identified, the observed development coincides with advances in QT and their improved accessibility. Thus, this study provides an empirical basis for further analyzing the development of QT-related art, the possible emergence of a "quantum art scene," and the reciprocal relationships between art and quantum technologies.

## Introduction

Discussions of a "quantum art scene" (Huffman, 2021) and indications of quantum's growing cultural influence (Rival Strategy, 2025; Selby, 2026; Yalcinkaya, 2025) have become increasingly prominent. While a broader artistic discourse explores the principles of quantum physics, which could potentially lead to changes in artistic perception and expression (Thomas, 2026) or the emergence of a new "quantum aesthetics" (Liu & Chen, 2026), the aforementioned discussions focus more specifically on quantum technologies (QT) based on these principles. Recent examples are art that conceptually explores QT without technically employing it (Kondratyeva, 2025; Mannone, 2022); art that deliberately uses QT to create artworks (Crippa et al., 2024; Miranda, 2024); or art that examines developments in QT and related questions of access, power relations, societal impacts, or philosophical aspects (Kroker & Cook, 2023; Zetti, 2024). While these approaches differ in their motivations and forms of engagement with QT, they may be relevant not only for the development of artistic practices, but also for how these technologies are perceived and discussed in society.

These cultural developments may have broader implications. In technology assessment, artistic engagement with QT is discussed not only in relation to the societal reception of these technologies, but also regarding its potential influence on technological development itself (Schmidt et al., 2025).

[*]Corresponding author: adrian.schmidt2@kit.edu

Relatedly, hermeneutic technology assessment examines visions and narratives of emerging technologies, as they provide insights into current expectations and interpretations of possible technological futures (Grunwald, 2020). This is particularly relevant for new and emerging technologies such as QT, whose future applications and societal implications remain uncertain. Artistic engagement with QT may therefore provide insights into such expectations and, through alternative visions, potentially influence future technological trajectories (Schmidt, under review).

Conversely, QT may also influence artistic practices. Drawing a parallel to photography, Huffman suggests that the specific technical capabilities of QT could enable new artistic practices and potentially establish quantum technologies as a distinct artistic medium (Huffman, 2021). From this perspective, "Quantum Art" could give rise to entirely new and different kinds of artworks, while the quantum phenomena underlying QT could themselves become part of their conceptual and aesthetic foundations (Shin, 2024). Such developments may also raise questions concerning access to QT, power relations, and the societal implications of their artistic use (Selby, 2026; Zetti, 2024).

While these perspectives point to potential developments in artistic practices and their broader societal relevance, it remains unclear how QT-related artistic engagement has actually evolved over time. This means examining whether the number of documented QT-related artworks has increased over time and how different forms of engagement with QT can be systematically distinguished and compared. The term "QT-related Art" is therefore used in the following to encompass different forms of artistic engagement with quantum technologies without requiring their actual technical use.

Initial qualitative and quantitative review studies have sought to address aspects of these questions. In quantum computer music, early overview studies (Miranda, 2024) and systematic reviews of scientific publications (Zheng & Yang, 2026) have been conducted, while qualitative studies highlight the increasing exchange between art and (quantum) science (Schmidt et al., 2026). However, a systematic, cross-media investigation of how many QT-related artworks have been documented, how their production has evolved over time, and which forms of engagement with QT they represent is still lacking. This is related to a missing framework to consistently find and classify those artworks, since current approaches partly overlap analytical dimensions (Solano, 2022), while others are rather medium-dependent (Itaboraí, 2023). This points to a research gap concerning the scale, temporal development, and different forms of QT-related artistic production, which this study seeks to address:

**Research question:** How has documented artistic engagement with quantum technologies evolved over time in terms of artwork production, and what temporal patterns can be identified regarding modes of QT-related engagement and artist participation?

This question is addressed through the systematic identification and mapping of a corpus of documented artworks, using a structured search and screening process informed systematic review methods, followed by longitudinal analysis. Temporal developments, modes of engagement with QT, and patterns of artist participation are examined quantitatively. Artworks are distinguished between engaging with QT conceptually, reusing QT-generated data, executing QT methods in classical simulation, or incorporating operations on quantum hardware. However, the study does not aim to establish artistic hierarchies or evaluate artworks based on their degree of technological involvement as a measure of artistic value. Instead, the study aims to provide a quantitative overview and contextual understanding of documented developments in QT-related artistic practices, thereby providing an empirical basis for understanding how QT-related artwork production has evolved. This may not only allow future research to identify potentially underexplored areas for research and artistic practice, but also to examine the potential cultural and societal implications of these developments.

# Theory

## Artistic practices and scientific-technological development

Artistic practices, scientific knowledge production, and technological innovations have historically been closely intertwined. While numerous artworks illustrate and explain scientific findings, others involve artists as active partners or critics in scientific and technological innovation processes (Wilson, 2001). Such interactions are also reflected in institutional collaborations, such as Experiments in Art and Technology (E.A.T.) which aimed to foster collaboration between artists and engineers for the benefit of society (Cateforis et al., 2018), or Arts at CERN, which provides artists with direct and open-ended access to fundamental scientific research (Arts at CERN, n.d.). For the purpose of this study, a broad definition of art is adopted, encompassing artistic practices such as visual arts, literature, film, music, performance, and digital media.

This relationship has led to different forms of mutual influence. On the one hand, technological developments have provided new subjects for artistic engagement and transformed artistic media and practices, as illustrated by photography and electronic music. On the other hand, artistic representations of future technologies can influence their societal perception and potentially stimulate technological development, as shown by Kirby for cinematic representations (Kirby, 2010). Artistic visions of the future can therefore contribute to societal negotiations about possible technological developments, which can be related to the concept of *sociotechnical imaginaries* (Jasanoff & Kim, 2015). Similarly, *hermeneutic technology assessment* examines present visions of technological futures and the societal expectations and evaluations reflected in them (Grunwald, 2020).

These reciprocal relationships are also relevant in the context of quantum technologies. Before turning to QT-related artworks, the following section therefore introduces QT and their foundations in quantum physics.

## Quantum physics and quantum technologies

Second-generation quantum technologies, hereafter referred to as "quantum technologies", exploit quantum-mechanical phenomena such as superposition and entanglement through the controlled preparation, manipulation and measurement of quantum states (Dowling & Milburn, 2003). Unlike conventional technologies that also rely on quantum physics, they require active control over individual quantum states. However, interactions with the environment can suppress the coherence required for these operations, making precise control very important.

Based on these principles, a wide variety of research and application areas have emerged, commonly categorized into three pillars. *Quantum computing* processes information encoded in qubits rather than classical bits (Benioff, 1980; Feynman, 1982); *quantum communication* uses quantum states for tasks including key distribution (Bennett & Brassard, 1984) and aims to connect quantum processors; *quantum sensing* uses quantum systems to measure quantities such as magnetic fields, acceleration and time with high sensitivity and precision (Degen et al., 2017). Despite their differing fields of application, these technologies share underlying quantum-mechanical principles and, in some cases, similar technological components.

Although the theoretical foundations were developed decades ago, access to the different quantum hardware remained largely limited to specialized research institutions. While for quantum computing early initiatives provided public access to quantum processors as early as 2013, broader accessibility emerged in the following years. In 2016, IBM introduced public cloud access to its quantum processors and, in 2017, released the open-source software development kit Qiskit in 2017 (Devitt, 2016; Javadi-Abhari et al., 2024). Another important milestone was reached in 2019, when Google demonstrated quantum computational advantage for a specific sampling task (Arute et al., 2019). In the following years, the field experienced substantial global growth in scientific publications, patent activity and commercial involvement (EPO & OECD, 2025; Lewis, Adam M et al., 2025). Today, many companies provide access to different quantum computing architectures, enabling different kinds of people to experiment with quantum hardware without operating their own systems (Schmidt et al., 2025).

Alongside the most heavily funded pillar of quantum computing, the other pillars have also made significant progress. Quantum communication, and especially quantum key distribution, which was proposed in 1984 (Bennett & Brassard, 1984), is nowadays being developed as a response to the threat of quantum computers hacking current encryption standards (Brazaola-Vicario et al., 2024). Quantum sensing has also made significant progress, including the development of established techniques and emerging quantum devices for commercial use (Degen et al., 2017). Therefore, even though quantum advantage for quantum computing has not yet been demonstrated in economically relevant applications (Preskill, 2025), and quantum sensing and quantum communication applications are not yet ready for the mass market (EPO & OECD, 2025), one can observe a clear trajectory towards this goal. This trajectory is not only evident in hardware developments.

## Artistic practices with quantum physics and quantum technologies

Quantum physics has long been adopted by artists in their creative processes. Examples date back to the early days of quantum physics; many consider Marcel Duchamp's work *3 Standard Stoppages*, 1913, which deals with probability, to be one of the first artworks of this kind (Thomas, 2018). Later, pieces such as *Quantum Metaphone* by Nick Herbert in the 1970s (Gómez, 2019) approached quantum physical phenomena artistically or are inspired by them. More recent examples include *Quantum Cloud*, 2000 by Antony Gormley (Gormley, 2000), or *Quantum Man*, 2007 by Julian Voss-Andreae (Voss-Andreae, 2011). The latter shows that also trained physicists engage artistically with quantum physics. Other examples of the strong relationship between science and art include artists like Paul Thomas, who collaborated closely with several scientists, including quantum physicists (Thomas, 2026). These varieties of relations between art and quantum science also account for quantum technologies.

Artistic engagement with QT has developed in various ways in recent years. On the one hand, there are works such as *Metaphase Soundmachine*, 2014 by Dmitriy Morozov (aka ::vtol::), in which sounds are produced by detecting existing environmental radiation with a Geiger counter, establishing interpretive connections to entanglement and quantum communication (Gómez, 2019). On the other hand, some artists also include quantum technologies in the artistic process, like an ion trap in *Ion Hole*, 2016, by Evelina Domnitch and Dmitry Gelfand, which is one approach to building quantum computers (Domnitch & Gelfand, 2016). In addition to these illustrative or experimental works on quantum technologies, there are also a variety of works in which QT or its elements are actively incorporated into and used within the artistic process. For example, Alain Lioret uses the fundamental principles of quantum technologies, such as qubit representations in many of his pieces (Lioret, 2016); Roman Lipski actively uses Qiskit in his works with Quantum Blur (Wootton & Pfaffhauser, 2021); and Akihiro Kubota explores real hardware in his Quantum Computer Art Studies (Kubota, 2025). The range of creators again extends from artists who collaborate with physicists to physicists who engage in artistic practice themselves. One particularly prominent example is theoretical physicist Libby Heaney, who has developed a wide variety of works ranging from more metaphorical approaches to pieces created with real quantum computer hardware (Hamilton, 2023; Heaney, 2019).

All of these practical works are accompanied by activities from various artistic, scientific, and technical organizations that seek to develop ideas for the diversity of current approaches. Notable examples include art-science collaborations, such as workshops on quantum computing and art (Schmidt, 2025), institutionalized programs (Carle, 2026), and QT seminars for artists or art students (Berlin University of Arts, n.d.). There are also global residency programs with numerous exchange opportunities, such as the Goethe-Institut's Studio Quantum (Goethe-Institut, n.d.), and international events that promote cultural diversity and different epistemologies, such as Quantum Crossroads during the International Year of Quantum Science and Technology (Brand et al., 2026). These are complemented by private-sector initiatives launched early on, such as those by IBM (Huffman, 2021), art contests (Center for Quantum Nanoscience, n.d.), and numerous exhibitions with different intentions (University of Bristol, 2017; Decaroli & Malinowski, 2022; Hamilton, 2023; Museum of Contemporary Art Tokyo, n.d.). While this emerging diversity demonstrates the rapid pace of development in the field, the exact trajectories of artistic practices within and involving quantum technologies remain unclear.

## Classification of QT-related artworks

The recent growth in QT-related artistic engagement led to the first studies assessing developments in QT and art. For example, Piispanen et al. work on quantum games included the assessment of quantum-related artworks - while the wider term “quantum games” is too wide for the paper here (Piispanen et al., 2025). For Quantum Computer Music, the work by Zheng & Yang is the first systematic review of Quantum Computer Music, resulting in 19 papers (Zheng & Yang, 2026), but being rather medium-dependent and focusing only on one of the QT. Other studies like *Quantum Art* by Alain Lioret (Lioret, 2016) are rather explorative or exemplary for certain directions with artistic engagement with QT, and thereby less extensive and systematic. Together those first studies show that there are various ways to assess QT and art developments.

More specifically, various dimensions of QT-related artistic works have been analyzed in varying depth. For instance, the application of quantum physics principles in artistic design is examined (Cattan et al., 2024); the potential applications of various media and feedback mechanisms are considered (Miranda, 2022); and the form of producing meaning of artistic works is differentiated. This can be broken down into three categories: *exploring* QT, *using* QT in the artistic process, and addressing the *societal embedding* of QT. These categories were derived from different artworks presented at a workshop organized by one of the authors (Schmidt, under review). Specific technical analyses also examine quantum technology, particularly quantum computers, as tools or instruments (Shin, 2024), analyze their concrete elements within the artistic process (Itaboraí, 2023), and examine the aspects of artistic works altered by QT from a more media-specific perspective (Lioret, 2016). These approaches are somewhat linked to a proposal by Enrique Solano to categorize artworks as *Quantum Inspired, Quantum Simulated, Quantum Sourced, Quantum Interactive, Quantum Feedback,* or *Quantum NFT* (Solano, 2022). Solano's ideas do not differentiate linearly, but are rather based on partially overlapping categories that can be understood as genres in their own right. Artworks are distinguished with an ideological influence from physics from those with the incorporation of experimental results or specific quantum technologies, up to responsive works or those in the cryptospace.

Taken together, current approaches differ in type and level of differentiation and highlight the absence of a unified classification framework for QT-related artworks. This makes it clear that "quantum art" is not yet defined. Therefore, a newly developed adaptation of these approaches is needed to systematically examine the diversity and development of these practices.

# Methods

To answer the research question, we combined different methodological approaches as no established classification standard for QT-related artworks existed yet. The identification of artworks was based on an extensive search across different databases, documented with reference to the search-reporting principles of PRISMA-S (Rethlefsen et al., 2021). The subsequent classification was informed by directed content analysis (Hsieh & Shannon, 2005), using theory-driven categories and available artwork information. Given the large number of identified artworks, LLM-supported screening and classification were employed, focusing on art-related texts (Garcia et al., 2020). Since LLM performance depends on the specific classification task, the results were checked against human-coded reference data (Halterman & Keith, 2026; Ziems et al., 2024).

The resulting overall procedure consisted of four steps: (1) *identification* via an extensive manual database search based on clear, inclusive criteria for quantum-related art; (2) LLM-supported, *screening* with regard to scope, QT relation, and date, using theory-driven, clear categories; (3) manual quality control to check *eligibility*; and (4) *analysis* of the results in light of the research questions.

## Identification

To identify as many existing artworks related to quantum technology as possible, we conducted a structured, iterative search across different databases and online sources using inclusive inclusion criteria. Based on these criteria, we developed and adapted search terms to the respective platforms. The search was conducted from July 7 to August 1, 2025. The search sources and strategies are documented in Table 1 and the appendix. The identified artworks were collected in a corpus database and cross-checked for potential gaps through additional searches and supplementary sources. Due to the diversity of online sources and the changing nature of web-based search results, the resulting corpus provides a broad mapping of QT-related artworks rather than a complete inventory. Focusing initially on quantum-related rather than exclusively QT-related artworks allowed for a more inclusive identification process, with artworks subsequently narrowed down to those with an explicit QT relation during screening.

The inclusion criteria were as follows: All artworks with some relation to "quantum" were included. Keywords such as "quantum art" or "quantum AND art" were included. All artifacts that were described or presented as artworks by their creators were counted as artworks, regardless of their medium, form, or degree of engagement with quantum technologies. This is based on the assumption that anyone can be considered an artist, including professional artists, scientists, and laypeople. However, tools and instruments were excluded unless they could be specifically attributed to a work of art; this also applied to quantum games. The time frame was limited to January 1, 2000, a pragmatic decision based on the initial definitions of second-generation QT in the early 2000s, through August 1, 2025, corresponding to the time of the search. We adopted a two-step process with the following input:

***Tab. 1*** *Artwork identification process*

| **Step 1: Artwork identification through database search** |
|---|
| - **Databases**: Search engines and bibliographic databases (Google, Yahoo, Scopus), Online art platforms (Google Arts & Culture, Artstor, Europeana, WikiArt), Galleries & Auction houses (Sotheby's, Artnet, Artsy), Research literature (Scopus, JSTOR), Museum databases (MoMA, MET, Tate, Belvedere, Pinakotheken, Louvre), Open access image archives (Rijksmuseum, Wellcome Collection, Digitales Bildarchiv Marburg)<br>- **Search string**: ("quantum art" OR "quantum and art" OR quantum artwork OR ("quantum-inspired" AND "art") OR "quantum aesthetics" OR ("quantum" AND "generative art") OR ("quantum" AND "visual art") OR ("quantum computing" AND art) OR (quantum computational aesthetics) OR ("quantum technologies" AND "art") OR ("quantum technology art") OR ("art" AND ("quantum superposition" OR "quantum entanglement" OR "quantum measurement" OR "quantum cryptography" OR "quantum algorithms" OR "quantum uncertainty")) OR "quantum computational interpretation of art" OR ("quantum art" AND ("exhibition" OR "installation" OR "artist")) OR "quantum music" OR "quantum computer music" OR (("quantum technology" OR "quantum technologies") AND music) OR (quantum composition)[1] |
| **Step 2: Artwork identification through other sources** |
| - Double-checking the corpus for missing artworks with ChatGPT and Gemini, using the same search string.<br>- Ensuring that the name of the artists' matched with the name of the artworks.<br>- Checking the authors' own database for additional artworks and artists, including scientific publications.<br>- Separate refinement step to check for duplicates and for spelling errors. |

The results were listed in the format number - artwork title - artist - primary link, checked for duplicates and used as input for the next step.

## Screening

The screening process combined two steps: narrowing down the identified artworks based on the inclusion criteria (individual artworks, clear connection to quantum technologies, appropriate year) and classifying the artworks into these categories as input for the subsequent analysis. For the large number of about 1,000 quantum-related artworks found, an LLM-supported classification method was chosen, since a content-intensive manual classification by two individuals would have required an enormous amount of time, including the resolution of conflicts by a third author. Together, this resulted in a multi-stage screening process using three classification prompts: (A) Scope, (B) QT relation, and (C) Date. Each was divided into identical, sequential individual phases: (I) drafting the prompt and iterative testing, (II) applying the prompt and recording the results, and (III) verifying the results.

[1] The keywords *without* quotation marks have been chosen as those keywords did not give any results when used with quotation marks.

Phase I - Prompt initialization and testing: The three prompts have been standardized following a scheme of introductory information, classification description, and output. The introductory paragraph defined how to execute the prompt and which artwork information was acceptable for classification. It referred to the stored link and other accessible sources, which were sources that loaded without a paywall. These sources contained identifiable artwork information, such as the artwork title and artist name. They also explicitly referenced the same artwork title or variants/translations (e.g., the same work with a subtitle) or allowed clear identification via catalog number, collection ID, etc. The classification was defined along the respective classification. The output always included the date of classification, the LLM version used, the sources used, and a comment explaining the classification.

Following this scheme, the prompts were successively developed and iteratively refined using a small test set of 20 artworks. When a first stable version was achieved, they were compared to a manual classification using a *reference subset*. To this end, 100 artworks were randomly selected from the entire dataset and first manually classified by two raters[2]. Discrepancies were resolved with the assistance of a third rater. The resulting reference classifications were then compared to the majority vote from three LLM-based classification runs using Cohen's kappa (Cohen, 1960). Cases without a clear majority were excluded from this comparison on both sides and reported separately. Following the interpretations by Landis and Koch, prompts with $\kappa > 0.8$ were reused directly, those with $0.6 < \kappa < 0.8$ improved as much as possible, and prompts with $\kappa < 0.6$ underwent fundamental changes (Landis & Koch, 1977).

Phase II - Application and recording of results: Prompts of the appropriate quality were manually entered into the latest ChatGPT model ChatGPT 5.2, 5.4, and 5.5 in "Thinking Mode". Ten artworks of the following type were entered each time: number - artwork title - artist - primary link. A new prompt window was used for each round, and memory functions were disabled. The results of each of the three runs were exported as tab-separated value (TSV) files and compiled into tables.

Phase III - Output analysis: The final results were checked for internal consistency and compared to the manual classification. To this end, Cohen's $\kappa$ was calculated for each pair of the three LLM classification runs. The arithmetic mean of these three values, denoted as $\langle\kappa\rangle$, was used to assess the stability of the classifications across repeated runs. Afterwards, another human-reference check was performed using the same reference subset ($\kappa$). The prompt was optimized where $\kappa$-values were below 0.6, and Phase I was restarted. Furthermore, unclear classifications of functioning sources, including unclear majority votes, and those from social media were manually reviewed. For artworks created before 2020, all respective artists were contacted and asked for a manual classification; 68.8% replied.

The three different prompts were refined into three final formats over the course of more than 20 substantial iterations. Their key points are outlined below.

*(A) Scope: How to clean the corpus*

The goal of the first prompt was to (a) check the corpus for incorrect artwork-title-and-artist combination and correct them, and (b) ensure that all corpus entries correspond to **single artworks**. For the first part, we checked for spelling errors by referring to the stored link and other accessible sources as outlined above. For the second part, the review determined whether the artwork documentation referred to a single, specific artwork, or to a collection, series, album, body of works, or an exhibition. If the documentation primarily described a collection of artworks, such as an album, or exhibition and did not clearly identify a single artwork with a distinct title and artist attribution, it was not aggregated across works, and the artwork was excluded as **non-single artwork**.

[2] For a more detailed description of the reference subset and the other steps, see the Appendix.

Some works of art yielded unclear or borderline classification results, which we attempted to resolve manually. Unclear "artwork title – artist name" combinations were manually reviewed, and if they remained unclear, they were removed. Several indistinguishable parts constituting a single artwork with a unique name - as with installations, for example - were treated as a single artwork. Multiple artworks from similar or the same collection were still treated as individual artworks. Ultimately, the artworks identified as "non-single artworks" were checked for possible individual artworks, thereby supplementing the list of artworks to be classified in the next step.

*(B) QT relation: How to classify the relation to quantum technologies*

The revised list of individual artworks was examined in terms of the QT relations afterwards. Building on the ideas of Enrique Solano (Solano, 2022), and the authors' preliminary work (Schmidt, under review), a categorical classification was developed to identify the possible types of relationships and uses of quantum technologies in the artistic process - meaning *if* and *how* quantum technologies have been used to create the artwork or if they were part of the artwork itself.

First of all, the distinction was made between artworks to be included and those to be excluded based on whether QT were addressed in any form within the artwork. All artworks engaging with QT in any way were included, while those with no explicit relationship to QT were excluded. Quantum technologies were defined as engineered systems and platforms that actively exploit controllable quantum states and phenomena, such as superposition, or entanglement, to enable computing, sensing, communication, and simulation capabilities. Classical simulations of such systems were included as well.

To answer the research question comprehensively and ensure the LLM could make this distinction precisely, the categories to be included and excluded were subdivided into subcategories before the classification. Regarding the non-QT artworks, we examined whether they were fundamentally quantum-related, such as artworks inspired by quantum physics, or if they had no quantum reference, thereby being "Classical". For a more precise evaluation, and to differentiate some artworks from those of included categories, the former category was subdivided further into "Quantum-Inspired" and "Quantum Measured." The latter refers to the active use of verified quantum (experimental) results, while the former includes conceptual works.

A similar, more nuanced distinction was made for the included QT-related artworks. Similar to the "Quantum-Inspired" category, we made a distinction between *passive* and *active* use of quantum technologies in the artistic process. The active use was then divided into simulated and real hardware, meaning that either classical hardware was used to mimic quantum processes, or real hardware was used to execute them. This made it possible to identify the extent to which QT serves as a concept ("QT-inspired"), are actively employed in the artistic process ("QT-generated" or "Simulated QT"), or passively form the basis for artistic engagement (QT-Sourced). The following categories resulted (full description in the appendix)[3]:

[3] In this way, the following aspects of Solano's and Schmidt's work are applied: On the one hand, Schmidt enables a distinction between work with QT (use) and work on QT (explore, societal embedding) (Schmidt, under review). On the other hand, following Solano, a distinction is made between active and passive use of QT, whereby a connection to hardware is also established here (Solano, 2022).

- **Classical Art (exclude)**: No mention of quantum physics or Quantum Technologies, or only the word "quantum" in the title or description, without any other quantum-related information.
- **Quantum-Inspired Art (exclude)**: Artworks that reference quantum physics, quantum phenomena, or quantum theory, but without a clear, explicit relation to quantum technologies.
- **Quantum-Measured Art (exclude)**: Artworks based on empirical data from quantum physics experiments that do not involve QT hardware.
- **Quantum Technologies-Inspired Art (include)**: Artworks that explicitly engage with quantum technologies, but do not use real QT-generated data and do not actively execute QT methods on simulated or real QT systems.
- **Quantum Technologies-Sourced Art (include)**: Artworks based on the passive use of data produced by a quantum technology platform, regardless of who produced the data and even if it was originally generated in a research context .
- **Simulated Quantum Technologies Art (include)**: Artworks actively created with QT methods, but simulated on classical technology. The simulated QT function itself must be the basis of artistic generation.
- **Quantum Technologies-Generated Art (include)**: Artworks where QT hardware actively performs operations that directly influence or generate the artwork.

Uncertainties also arose during this classification step and were resolved manually whenever possible. Initially, we relied solely on the artists' descriptions. This means that we did not verify the use or non-use of certain quantum technologies if this information was not clearly indicated in the artwork description. For example, ion traps, which can serve as building blocks of quantum computers (included) *or* represent a quantum physics experiment (excluded), were only categorized as quantum technology if this was explicitly stated in the artwork description[4]. If the relationship to quantum technology was unclear and multiple categories were possible, the most appropriate one was selected. If this decision could not be made, the category with the clearest use of quantum technology was chosen. This meant that one of the later categories in the list above was prioritized over an earlier one. However, these cases do not imply any kind of hierarchy, as the study is not about defining "what has the strongest QT connection". These are predominant design choices[5]. Regarding QT-Generated Art specifically, the use of any QT hardware to create QT-related art is sufficient for this category. This means single physical qubits suffice to qualify as QT; no gates, circuits, or algorithms need be executed. All artworks whose QT relation could be classified were subsequently moved on to the next classification step.

---

[4] This is because the boundaries between QT and non-QT are fluid and more a matter of definition. Therefore, the self-descriptions provided by the artists or participating physicists offer the clearest distinction. For instance, *Ion Hole,* 2016 by Evelina Domnitch and Dmitry Gelfand was included in the list of artworks with QT relation despite having only few artistic references to quantum computing because the expanded description clearly states that ion traps, used in the artwork, are a potential foundation for quantum technologies (Domnitch & Gelfand, 2016).

[5] This decision could be made differently in other studies. For example, QT-sourced art directly uses the output of quantum technologies to create art. Therefore, it is not about programming something classically that uses quantum principles in the background (as with quantum-generated art). Those pieces could also be prioritized in the design choices in the future. However, since these pieces currently focus more on using sensor data or single measurements than on a "fully embedded quantum process," we proceeded as we did.

*(C) Date: How the classification of the dates has been done*

To examine trends over time, all works of art with a clear QT relationship were classified by their identifiable years of origin. Since the year of creation could not be identified for all artworks, the first public online mention and the first identifiable public presentation were also determined, which was appropriate given that the search was based on online databases and some links referenced past exhibitions. This resulted in the following three date classifications:

- **Year of completion**: This is the year in which the artwork itself was completed, created, produced, realized, or dated as a work, when explicitly assigned in the artwork source.
- **Year of first public presentation**: This is the earliest year in which the artwork was publicly presented, exhibited, or otherwise made public in a documented way.
- **Year of first online mention**: This is the earliest year in which the artwork itself is documented online in an accessible web source, verified via artwork-title-and-artist combination.

Subsequently, the first year that could be clearly identified by a majority vote among the three classification runs was used for further analysis, always starting with the year of completion if existing.

### Eligibility check

This comprehensive classification procedure was followed by a reexamination of the artworks' eligibility based on the aforementioned criteria. After checking for duplicates, which arose from corrections to the artwork-title-and-artist combination, individual artworks were reviewed for correct classification. This was particularly important for works created before 2020, due to the field's later-recognized origins during this period. We therefore contacted all artists from this period and asked them to assess their works based on our categories. Any suggested adjustments were then cross-referenced with the manual classifications we performed in parallel. Works additionally identified by the artists were added to the corpus.

### Analysis

The subsequent analysis was conducted on the entire dataset of individual artworks for which all three classification steps could be performed and for which a unique QT relationship existed. The *First year*, meaning the year of creation - or the following year(s), if unavailable - had to fall between January 2000 and August 2025. This analysis primarily focused on the artworks, their QT relationships, and their respective years of creation because it was limited to artworks accessible online. Thus, a comprehensive search of artworks by artists was not possible. This aspect was considered only to highlight potential overarching trends.

## Results

The results were analyzed based on artworks that could be unambiguously classified during the screening process, following the inclusion criteria. These results are presented below, along with the corresponding quality-assurance classification metrics. The analysis was conducted for individual categories and combinations of categories. It also included an examination of overarching insights derived from artworks by the most frequently represented artists.

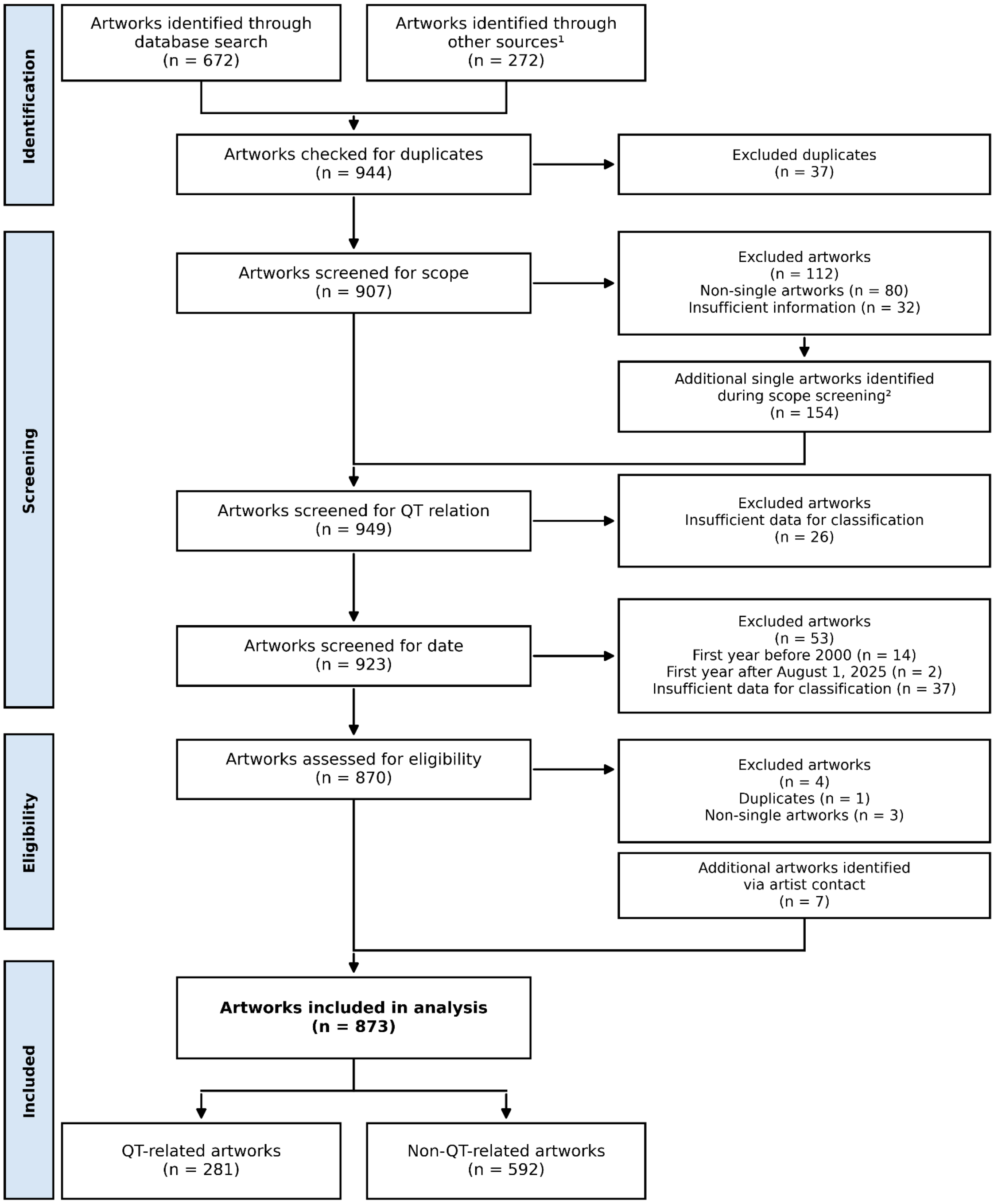

***Fig. 1*** *PRISMA flow chart for identification and selection of artworks. [1]Other sources include AI-assisted database refinement and manually compiled lists of the authors. [2]Excluded non-single artworks which represent a series or collection were checked for missing single artworks and added to the screening process.*

The final corpus consists of 873 artworks, which is based on an initial identified set of 944 artworks. Of these, 232 were removed due to duplicates, missing information, or failure to meet inclusion criteria, while 161 additional artworks were added. The exact breakdown is shown in Figure 1.

## Validation of the AI-based classification

The reduction from the initial set to the final corpus was based on a three-step, LLM-assisted screening process overseen by clear metrics calculations. Each step in Figure 1 was first carried out and deemed satisfactory when the corresponding quality assurance metrics $\kappa$ and $\langle\kappa\rangle$ were greater than 0.6. According to the methods, the following were calculated for all three screening steps (Scope, QT relation, and Date): (i) prompt-development performance $\kappa$ for the reference subset, (ii) full-dataset internal consistency $\langle\kappa\rangle$, and (iii) a final human reference check ($\kappa$) using the reference subset. The following is a summary of the results. For a more detailed description, including the respective used LLM model, see appendix.

After optimizing the prompts through iterations, all quality criteria met the necessary thresholds for application and acceptance of the screening results. Kappa values between 0.63 and 0.72 were achieved for the classification of the "scope" category when artworks found via social media links, which could not be opened or classified by the LLM, were excluded initially and classified manually later. This approach was also applied to the other two categories. In the case of the QT relation, Kappa values slightly below 0.6 were initially achieved despite multiple iterations of prompt refinement. However, it was determined that the cause was not only the large number of categories and the greater potential for differing classifications, but also a large number of borderline decisions between the excluded categories "Classical Art" and "Quantum-Inspired Art." Therefore, the quality metrics were evaluated in relation to the central study results. Considering artworks with exclusively included QT relations and calculating Kappa values to distinguish fundamental inclusion or exclusion consistently produced Kappa values above 0.6. In other words, artworks with some relation to quantum technologies were consistently classified as such, as well as their exact classification within QT-related categories was also mostly correct. For the final assessment of the "Date" classification, which focused on the first identifiable year ("First Year"), significant agreement was achieved, with Kappa values of 0.8 or higher. Overall, it was found that, although the quality of the QT classification was somewhat lower due to its qualitative basis, it still yielded good enough results to allow for further use in the research process, as with scope and date.

## Characteristics of the final dataset

The final dataset of artworks used for the analysis consists of 873 artworks with a clear *Scope* (single artwork), clear *QT relation* (all artworks with clear majority vote for one category) and a *First year* between January 2000 and August 2025. As seen in Figure A2 (appendix) those years are slightly shifted to younger dates, since the artworks might have been produced earlier than the first year of public presentation or online mention one could find online. For the QT relations the focus lies on artworks of included categories, meaning the artworks with clear relations to quantum technologies, but keep the non-QT-related artworks for comparison.

The first information about artistic practices related to quantum technologies can be found in the total numbers for QT-related artworks. As seen in Table 2, in our corpus are 281 artworks (32.2%) with date and clear QT relation. Of those included ones 17.8% are QT-generated, 40.9% Simulated QT, 9.6% QT-Sourced, and 31.7% QT-inspired, showing that all categories are filled with significant numbers.

***Tab. 2*** *Total number of artworks per category of QT relation.*

| QT relation | All | Quantum-related | QT-related |
|---|---|---|---|
| Classical Art | 245 | | |
| Quantum-Inspired Art | 325 | 325 | |
| Quantum Measured Art | 22 | 22 | |
| Quantum Technologies-Inspired Art | 89 | 89 | 89 |
| Quantum Technologies-Sourced Art | 27 | 27 | 27 |
| Simulated Quantum Technologies Art | 115 | 115 | 115 |
| Quantum Technologies-Generated Art | 50 | 50 | 50 |
| **Total Sum** | **873** | **628** | **281** |

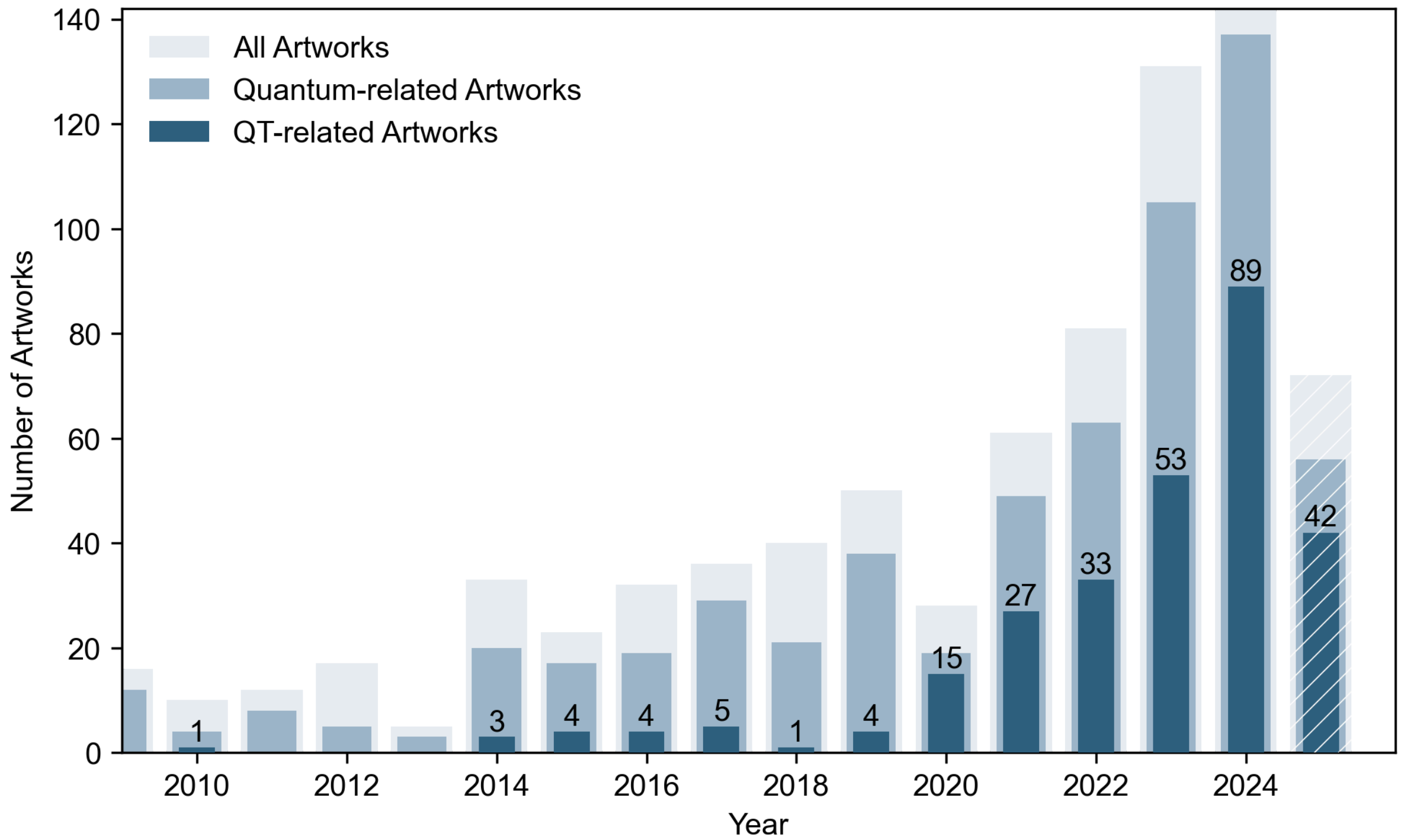

***Fig. 2*** *Snapshot of the corpus between 2010 and 2025. Shown are all the artworks included in the analysis, the proportion of those that have a quantum connection, and, of those, the portion that is QT-related. The numbers from 2025 are partially due to the cutoff date in August.*

To answer the question how the artistic engagement with quantum technologies has evolved, the temporal development of QT-related artworks was analyzed. As seen in Figure 2, the total number of artworks, including classical artworks, is steadily increasing until 2024[6]. Since classical artworks have not been the focus of the study and the search process, those are incomplete and not shown here in detail. For quantum-related artworks, one can find pieces from 2000 on in the database, e.g. *Quantum Cloud*, 2000 by Antony Gormley being one of the first (Gormley, 2000). Since many similar works by Antony Gormley have been found, there is a clear peak in 2000 (see Figure A1 in appendix). This peak is followed by rather stable low numbers, with quantum-related artworks increasing since 2014, showing some increasing interest in the topic in the last about ten years.

[6] When the number of artworks is extrapolated linearly after August 1st 2025, the number of artworks in 2025 would reach numbers slightly below those of 2024 (72 QT-related artworks).

Interestingly, the earliest QT-related artwork identified in this corpus can be found in 2010, followed by a few pieces between 2014 and 2017. After this first phase of QT-related artwork development, including some dip in 2018, the number of QT-related artworks significantly increased from 2020 onwards. This gives rise to a two phase development for QT-related artworks between 2014 and 2018 with low numbers and between 2019 and 2025 with steadily increasing numbers.

### Temporal distribution of QT-related artworks

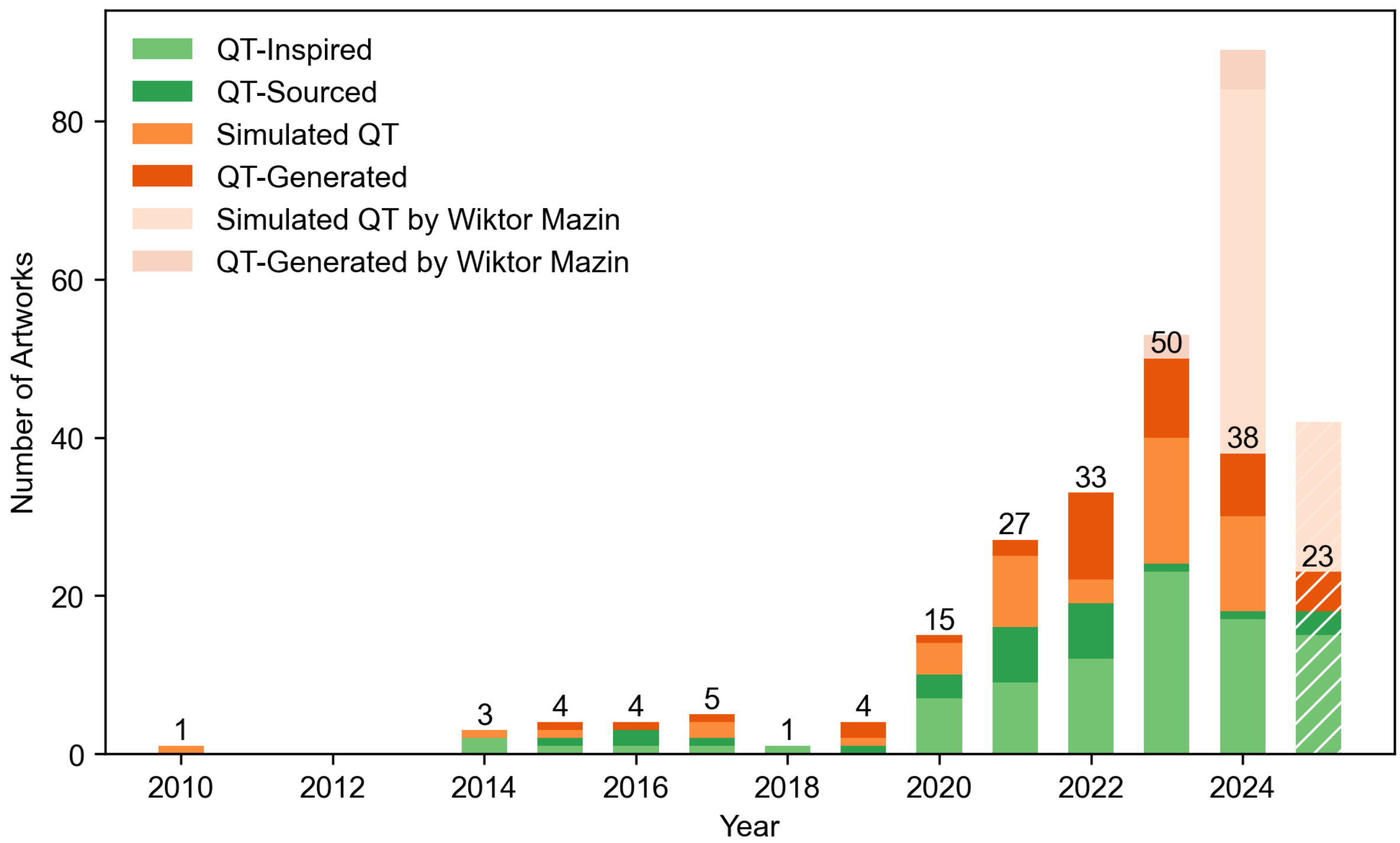

***Fig. 3*** *QT relation of included artworks per year. The artworks by Wiktor Mazin are added in semi-transparent colors. The numbers from 2025 are partially due to the cutoff date in August.*

To further assess the temporal development of QT-related artworks, the annual numbers in Figure 2 are differentiated by QT relation in Figure 3. The categories show varying temporal trends and shares. We therefore distinguish an initial phase (before 2019) and a second phase (2019 onwards), followed by a sensitivity analysis of Wiktor Mazin's works. These are highlighted in Figure 3 and account for 26.0% of the QT-related artworks in the corpus.

The first phase before 2019 marks the starting point for QT-related artwork creation in our corpus. The earliest piece is *Evalo 20M* (2010) by Alain Lioret, classified as Simulated QT Art. In that piece, Lioret manipulates a qubit representation as one of the building blocks of quantum technologies with continuous time quantum walk computation (Lioret, 2016). The next pieces appear four years later: *Quantum Swan* (2014), again by Alain Lioret (Lioret, 2016), *Metaphase Soundmachine* (2014) by Dmitriy Morozov (::vtol::), and *Observation collapse* (2014) by Ruixian Ma & Mercedes Gimeno-Segovia (Walker & von Ompteda, 2016), classified as Simulated QT, and two-times QT-inspired, respectively. The first QT-generated artwork appears with *Talking to an electron* in 2015.

In this piece, Paul Thomas collaborated with Andrea Morello and Juan Pablo Dehollain to actively manipulate phosphorus-doped spin qubits, by transmitting readings from Richard P. Feynman's 1982 paper (Feynman, 1982) to the qubit in the form of microwave signals and later visualizing the qubit responses artistically (Thomas, 2018). Other works of that period include *CLOUD* (2015) an early piece by Libby Heaney that focuses rather *on* QT (Heaney, 2019); and *Superposition* (2016) by Alexis Kirke, who used an adiabatic quantum computer to make music (Kirke, 2019). These examples illustrate the diversity of early QT-related artworks.

The second phase since 2019 accounts for most QT-related artworks in the corpus. QT-Inspired artworks broadly follow the overall development, with a steady increase in absolute numbers until 2024. The annual share of QT-inspired artworks ranges between 19.1 and 46.7 %. An example is Maria Mannone's drawing *Quantum Sphinx* (2021) displaying the quantum circuit of the Grover search algorithm (Mannone, 2021). The number of QT-Sourced artworks instead peaks in 2021 and 2022 (7 artworks each) followed by some decline. Examples include Refik Anadol's *Quantum Memories Nature Studies* (2021), which uses quantum computer outputs as input for the further artistic process, and *Quantum Sound* (2019), in which Spencer Topel, together with Kyle Serniak, Luke Burkhart, and Florian Carle explored a similar approach musically (Topel et al., 2022).

By contrast, the number of Simulated QT and QT-Generated artworks combined, representing the active use of simulated or real quantum hardware, increases after 2019, with a peak in 2024 (together 71/89). Simulated QT already accounts for 9 of 27 artworks in 2021, while QT-Generated artworks increase to 11 of 33 in 2022. Examples of Simulated Qt are *Quantum Blur* (2021) by Roman Lipski who implemented algorithms on quantum computer simulators, and *Fluidity I* (2023) or *Equations of Coltrane* (2023) by Scott Oshiro and Omar Costa Hamido combining live acoustic instruments with a quantum synthesizer, and quantum computer improvisation systems. QT-Generated artworks are particularly represented by works of Libby Heaney, such as the iteratively developed *Ent- (Non Earthly Delights)* (2024), which explores the transformative potential of the "forthcoming quantum future" partly through the integration of real quantum hardware. Overall, the second phase of the QT-related artworks development starts with a larger share of QT-inspired pieces or the passive use of real QT data, while the share of artworks actively engaging with QT is higher since 2021.

As a sensitivity check we explicitly analyzed Wiktor Mazin's 73 artworks, since they represent 5,7% of the corpus in 2023, 57,3% in 2024 and 45,2% in 2025. Excluding Mazin's artworks changes the temporal distribution considerably, with the highest annual number of QT-related artworks shifting from 2024 to 2023. Of his works, 89.2% have been categorized as Simulated QT, which explains why this category is clearly dominating the overall distribution in 2024 and 2025. In these artworks, Mazin creates quantum fractals of various structures and colors by converting quantum-derived complex numbers into fractal-generating functions, clearly showing that those are either created with simulated quantum computers or sometimes even real hardware (Mazin, 2026).

Finally, the excluded quantum-related categories provide additional insights. Quantum-Measured artworks remain rare, with zero to four works per year. Conversely, the number of quantum-inspired artworks have increased continuously, with a relatively large share between 14.3 and 35.8 %. Overall, these numbers demonstrate not only that the number of quantum-related artworks has increased significantly over the past 15 years but also that passive and active engagement with quantum technologies has grown substantially during this period.

## Involved artists in QT-related artwork creation

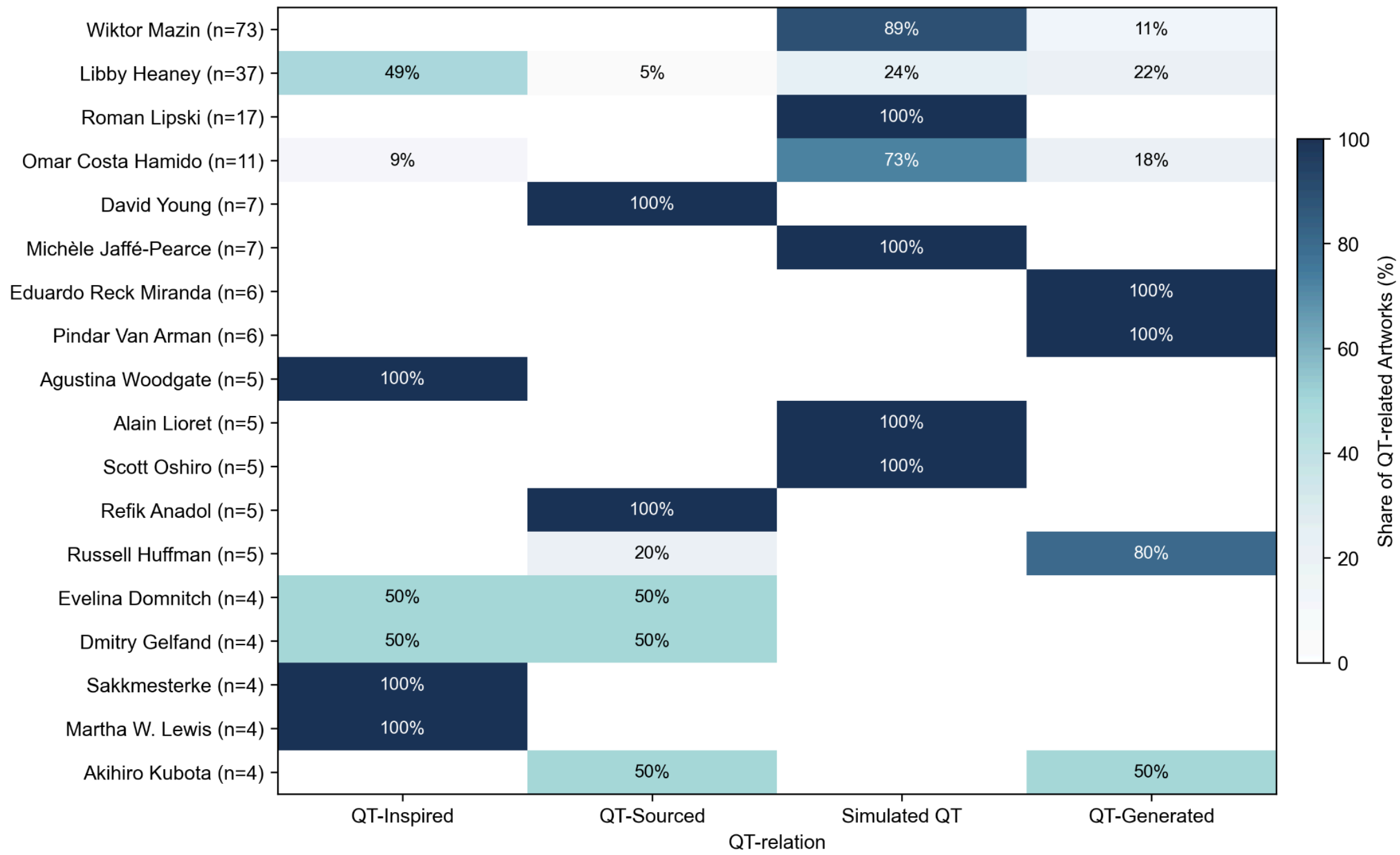

***Fig. 4*** *Share of artists' artworks for the different QT-related categories of artists with at least 4 QT-related artworks. n is the total number of included QT-related artworks of the respective artists. Their total number in the database may vary if all categories are considered.*

The dominance of certain artists in different periods (e.g. Alain Lioret in the early years or Wiktor Mazin more recently) raises the question of how individual artists contribute to the observed temporal developments and the distribution of QT relations. Since, unlike for scientific publications, artwork descriptions rarely allow for tracing individual artistic developments, we instead examined the distribution of QT relations among artists with multiple documented artworks, as well as the year of their first documented QT-related artwork in our corpus.

Of the 191 identifiable individual artists who have been involved in the creation of at least one QT-related artwork, we found 43 artists with more than one QT-related artworks of which 28 created more than two.[7] The top 17 of those artists can be seen in Figure 4, showing that the top 4 artists account for about 50% of the QT-related artworks, including Wiktor Mazin (26.0%), Libby Heaney (13.2%), Roman Lipski (6.0%) and Omar Costa Hamido (OCH) (3.9%). In terms of temporal development, many artists created their first QT-related artwork in the year 2021 (43 artists, or 22.5%) or in 2023 (41 artists, or 21.5%), showing that 77.0% of the artists entered the field in 2021 or later.

The distribution of QT relations varies considerably between individual artists. While three of the top four artists created artworks in at least two different QT categories, five of the top eight focused on a single QT relation. Among these eight artists, Simulated QT and QT-Generated artworks account for 136 of 164 works (82.9%), or 63 of 91 (69.2%) without Wiktor Mazin. In contrast, among artists with five or fewer QT-related artworks in the corpus, only 86 of 253 works (34.0%) fall into these categories, with the majority being QT-Sourced or QT-Inspired[8]. Overall, artists with more documented QT-related artworks tend to focus more on the active use of QT methods or hardware in their artistic processes.

[7] 147 of the 191 have collaborated with others in the creative process at least some of the time, while 24 have always worked as duos (see Evelina Domnitch and Dmitry Gelfand in Figure 4). In the following analysis, each artist from a duo who could be identified as individual people is counted as a single artist.

## Discussion

This study produced a variety of quantitative results that can be correlated with known developments in QT and QT-related art. The key findings are presented below. As the first study of its kind, it introduces methodological innovations and, consequently, limitations. These limitations are discussed below, along with a critical examination of the use of LLMs and possibilities for future research.

### Key results

The earliest QT-related artwork in our corpus dates back to 2010, with Evalo 20M by Alain Lioret, who focused primarily on qubit representations and thus on the simulation of fundamental elements of quantum technologies. This approach, closely linked to the basic principles of the technology, is also reflected in the chronologically following works, while from the mid-2010s onwards more direct artistic engagements with quantum technologies such as *Talking to an electron* (2015) by Paul Thomas, become apparent. Until 2019, this development remained relatively stable on a small scale involving only a few artists, indicating a distinct first phase of QT-related art.

The development of this first phase may be understood in the context of broader technological, scientific, and cultural developments. It becomes apparent that the first QT-related artworks emerged at a time of increasing research and development attention towards QT, when laboratories were increasingly exploring possibilities for implementing QT - particularly quantum computers - and when strategic support for QT was expanding, particularly through larger national and international funding initiatives (Knight & Walmsley, 2019; Riedel et al., 2019). Subsequently, early formats of exchange between science and art in this field were initiated, for example through collaborations at Yale that led to artworks and exhibitions from 2016 onwards (Carle, 2026). Early art-theoretical contributions also emerged, particularly in the field of quantum computer music (Miranda, 2017). It is therefore not surprising that many of these first artworks were of rather exploratory nature, created by artists who were themselves involved in the technology (cf. Libby Heaney) or who were in close exchange with those working in the field (cf. Paul Thomas). The first phase of QT-related art development was thus largely characterised by a scientific and exploratory approach.

From 2019 onwards, this first phase was followed by a distinct accelerated second phase. Of the 281 QT-related artworks, 263 (93.6%) were dated 2019 or later, while the number of participating artists also increased substantially, particularly from 2021 onwards. This increase was simultaneously accompanied by a differentiation and more extensive artistic engagement with quantum technologies, for example by Refik Anadol in *Quantum Memories Nature Studies* (2021). Besides this rather passive use of QT outputs, which represents a decreasing share of the corpus over the years, many artists show an increasing active engagement with quantum technologies, accounting for 157 of the artworks since 2019. At the same time, however, artworks also emerged during this period that reflect on the development of QT itself, particularly those by Libby Heaney. In a number of her works, such as the iteratively further developed *Ent- (Non Earthly Delights)* (2024), she reflects on the transformative potential of the "forthcoming quantum future", partly also through the integration of real quantum hardware. Although many artists thus continue to engage with QT conceptually, the second phase shows a substantial increase in artworks actively using QT in the artistic practice, with the relative distribution strongly influenced by a few highly productive artists.

[8] The total number of 417 represents the number of individual artists involved in creating QT-related artworks, including several collaborations, and therefore exceeds the number of 281 unique QT-related artworks.

The increase in documented QT-related artworks coincided with broader access to quantum technologies. This includes a steep increase in scientific publications (Lewis, Adam M et al., 2025), and increased access to quantum computing hardware and software (Devitt, 2016; Javadi-Abhari et al., 2024). Notably, IBM's cloud-based quantum computing platform, available since 2016, and Qiskit, released in 2017, have enabled artists to experiment with simulated and real quantum hardware (Huffman, 2021). Furthermore, Google's quantum advantage demonstration (Arute et al., 2019) marked an important milestone in quantum computer development and led to significant attention. accompanied also by a number of artistic and cultural events, such as residencies (Goethe-Institut, n.d.) or exhibitions (Decaroli & Malinowski, 2022; Hamilton, 2023). Thus, the accelerated development of QT-related artworks coincides with the broader expansion of quantum technologies, particular quantum computing.

Beyond this purely quantitative increase, the growing availability of QT hardware, software, and data may also help explain changes in the forms of artistic engagement with QT. QT-Sourced artworks, for example, incorporate outputs from quantum hardware into artistic processes, including data generated by the more recent Google Quantum AI and IBM Quantum processors. With the subsequent expansion of cloud access and programming tools, the increasing number of Simulated QT and QT-Generated artworks also reflects more active forms of technological integration into artistic practices. IBM's tools played a prominent role in this development, as illustrated by *Quantum Blur*, developed in cooperation with IBM and used by Roman Lipski to create audiovisual artworks (Wootton & Pfaffhauser, 2021). At the same time, independently developed tools emerged, such as the QAC Toolkit by Omar Costa Hamido for quantum computer-based music composition (Hamido, 2022). The expanding availability of real quantum hardware is also reflected in the 50 QT-Generated artworks in our corpus, including works using hardware from other providers, such as Quandela and IQM, whose cloud services became available in 2022 and 2024, respectively. At the same time, our data indicate that the active integration of QT software and hardware into the creative process is associated with a higher output of QT-related artworks among the artists involved (cf. Mazin, Heaney, Lipski, and Hamido). However, it remains unclear whether this reflects an accelerated creative process, a stronger focus on QT-related art, or other factors.

## Limitations

Despite the extensive search, the corpus may contain selection and documentation biases. The focus on English-language search terms, limited online availability of older artworks, and changes in online documentation may have affected identified temporal developments. Furthermore, the observation period ends in August 2025, meaning that more recent developments are not covered and the data for 2025 represent only a partial year.  Additionally, supplementary searches based on identified artists and the varying responsiveness of the artists contacted may have influenced the representation of individual creators. The treatment of series and individual artworks may have affected the observed numbers and their concentration among artists. Thus, the apparent increase in QT-related artworks may reflect increasing artistic production, greater online visibility, and improved documentation. Therefore, the corpus supports the analysis of documented trends rather than an estimate of the total number of QT-related artworks produced during the observation period.

The exploratory nature of the research can consequently lead to various qualitative classification errors. One issue is the lack of distinction between artists, participating physicists, and other actors in this study. This distinction was made based on the artwork descriptions, however, future studies could refine this distinction to improve accuracy regarding the number of involved actors. Additionally, the small number of QT-related artworks to date has resulted in few classification schemes and a lack of experience with borderline cases between categories. This may have led to misclassifications, particularly in borderline cases. Future work could address this issue to reduce variation.

Other limitations concern the classification framework and its validation. Using the same human-coded reference subset for prompt initialization and testing as well as output analysis may have led to an overestimation of classification performance. Furthermore, the operational distinction between quantum physics and QT may have affected the inclusion of artworks. While our definition includes artworks involving or referring to individual qubits, stricter definitions could lead to fewer included works. Experimental apparatus such as ion traps can also be used for both quantum physics experiments and QT, potentially leading to misclassifications when their intended use is not clearly stated in the artwork descriptions.

### Critical reflection on the use of LLM-assisted classifications

The decision to use LLM-assisted classification was based on the large number of artworks. Although the metrics show that it was used carefully and successfully, the authors do not want to imply that it is the only or correct method for conducting such a study. They have reflected on the disadvantages of this method, including the use of resources and dependency on non-human actors. This dependency may have resulted in a lack of personal engagement with the artworks, a problem that was tried to address through many manual cross-checks. Therefore, while the authors acknowledge that current AI system developments may significantly improve the quality of future LLM-based classifications, they also encourage other researchers to engage with artworks manually in future studies.

## Outlook

The significant increase in the number and variety of QT relationships among artworks underscores the need for regular updates to this and similar studies. For example, the International Year of Quantum Science and Technology in 2025 may have further stimulated artistic engagement with QT. Furthermore, the large share of individual artists such as Wiktor Mazin in recent years highlights the importance of monitoring developments in artist participation. Repeatedly searching for and classifying additional artworks using the same methods might be feasible. More standardized, API-based assessments using several different LLMs might also be applicable. At the same time, more traditional approaches seem possible. Future studies could examine individual artworks and their QT relation in greater depth, particularly through direct involvement of the artists themselves. This could help find new ways to create artworks with, about, or in relation to quantum technologies. Since we do not view our work as the standardization or automation of QT-related art, we strongly advocate for further hands-on research on quantum technology-related artworks by artists, quantum technology researchers, and other interested parties, especially in collaboration.

In light of this, further theoretical work on classifying QT-related artworks is necessary. Such work could address potential problems, such as unintended hierarchies in the use of QT and ambiguities in the categories and terminology. It is important to note that the distinction between quantum physics and QT is analytical and should not be interpreted as a division of artistic practices, especially considering the close relationship between QT and fundamental quantum research. The proposed categories are not intended to establish artistic hierarchies or guide artists to create the "most QT-related" artworks. Rather, they are intended to provide a framework for analyzing and reflecting on current developments. Thus, the categorization approach chosen in this study should be critically examined and expanded where appropriate. Future research could further differentiate the role of quantum operations within artistic processes and examine how quantum and classical computational elements are combined in the creation of artworks. Furthermore, as QT-related art continues to evolve, distinguishing between conceptual references and clearly documented technological involvement may become more difficult. These cases could be examined through a closer analysis of artwork documentation and dialogue between artists and QT researchers.

Additionally, it seems conceivable to analyze additional dimensions that have not yet been considered here. For instance, future studies could more narrowly define the concept of art. The medium of the artworks could be examined, or different types of quantum technologies could be distinguished. One could also investigate the extent to which artworks based on quantum computers dominate QT-related art. More detailed technical analyses could also distinguish how individual qubits, quantum gates, or entire algorithms are used within artistic processes.

Finally, future research could complement the quantitative analysis presented here by examining the narratives and visions associated with QT-related artworks and their development over time. This could involve not only investigating how developments in QT influence artistic practices, but also how artistic works reflect or potentially influence societal expectations and visions of QT. Drawing on hermeneutic technology assessment (Grunwald, 2020)grun, such research could provide insights into present interpretations of possible and alternative QT futures. Finally, future research could complement the quantitative analysis presented here by examining how the narratives and visions associated with QT-related artworks have developed across the identified phases.

## Conclusion

Our corpus clearly documents the growth of QT-related artistic practices over the past 15 years. Beginning with the first identified artwork in 2010, it shows a steep increase in artistic production since 2019 leading to 281 QT-related artworks by August 2025. At the same time, the number of artists involved remains relatively small, with a few artists accounting for a considerable share of the artworks, particularly those using QT actively in their artistic process. Still, despite the increasing variety of artistic approaches, no single trajectory in artistic practices can be identified.

To investigate these developments, we combined a structured search with a theory-guided classification approach and LLM-assisted screening, checked against human-coded reference data. This provides a basis for future studies investigating QT-related artistic practices. However, we do not aim to establish any hierarchy between the different forms of QT relations and encourage other researchers to continue engaging with individual artworks and their use of QT.

The increase in documented artworks coincided with the broader development of QT and the growing availability of quantum software and hardware. While these developments may have supported new artistic practices, our data cannot establish causal influences. Similarly, our findings align with discussions of an emerging "quantum art scene," though they do not demonstrate the existence of a distinct artistic medium or an established scene. How further developments in quantum technologies may influence artistic practices, and to what extent these artistic practices might, in turn, influence technological development and broader societal expectations about QT, remain to be investigated in future studies.

## References

Arts at CERN. (n.d.). *Artistic residencies*. Arts at CERN. Retrieved 15 October 2026, from https://arts.cern/artistic-residencies/

Arute, F., Arya, K., Babbush, R., Bacon, D., Bardin, J. C., Barends, R., Biswas, R., Boixo, S., Brandao, F. G. S. L., Buell, D. A., Burkett, B., Chen, Y., Chen, Z., Chiaro, B., Collins, R., Courtney, W., Dunsworth, A., Farhi, E., Foxen, B., … Martinis, J. M. (2019). Quantum supremacy using a programmable superconducting processor. *Nature*, *574*(7779), 505–510. https://doi.org/10.1038/s41586-019-1666-5

Benioff, P. (1980). The computer as a physical system: A microscopic quantum mechanical Hamiltonian model of computers as represented by Turing machines. *Journal of Statistical Physics*, *22*(5), 563–591. https://doi.org/10.1007/BF01011339

Bennett, C. H., & Brassard, G. (1984). Quantum cryptography: Public key distribution and coin tossing. *Theoretical Computer Science*, *560*, 7–11. https://doi.org/10.1016/j.tcs.2014.05.025

Berlin University of Arts. (n.d.). *Quantum Futures – Designing the Futures of Quantum Technologies*. Universität Der Künste Berlin. Retrieved 8 October 2026, from https://www.udk-berlin.de/studium/studium-generale/kurse-courses/seminare-kulturwissenschaft-sose-2026/quantum-futures-designing-the-futures-of-quantum-technologies/

Brand, J., Barder, J., Puah, X. Y., Vishveshwara, S., Costa Hamido, O., Andrews, H., Gadelha, D., Barreto Lemos, G., Choi, H. J., & Okamoto, T. (2026). Quantum Crossroads: A vision for art, culture and quantum science and technology. *Zenodo*. https://doi.org/10.5281/zenodo.18489718

Brazaola-Vicario, A., Ruiz, A., Lage, O., Jacob, E., & Astorga, J. (2024). Quantum key distribution: A survey on current vulnerability trends and potential implementation risks. *Optics Continuum*, *3*(8), 1438–1460. https://doi.org/10.1364/OPTCON.530352

Carle, F. (2026). Rigorously undisciplined. *Nature Physics*, *22*(4), 499–499. https://doi.org/10.1038/s41567-026-03250-2

Cateforis, D., Duval, S., & Steiner, S. (2018). *Hybrid Practices: Art in Collaboration with Science and Technology in the Long 1960s—PhilPapers*. University of California Press. https://philpapers.org/rec/CATHPA

Cattan, G., Duś, K., Kusmia, S., & Stopa, T. (2024). Art makes quantum intuitive. *Frontiers in Quantum Science and Technology*, *3*. https://doi.org/10.3389/frqst.2024.1397130

Center for Quantum Nanoscience. (n.d.). *The World of Quantum*. Qns.Science. Retrieved 8 October 2026, from https://qns.science/art2019/

Cohen, J. (1960). A Coefficient of Agreement for Nominal Scales. *Educational and Psychological Measurement*, *20*(1), 37–46. https://doi.org/10.1177/001316446002000104

Crippa, A., Chai, Y., Hamido, O. C., Itaborai, P., & Jansen, K. (2024). *Quantum computing inspired paintings: Reinterpreting classical masterpieces* (arXiv:2411.09549). arXiv. https://doi.org/10.48550/arXiv.2411.09549

Decaroli, C., & Malinowski, M. (2022). "Do you speak quantum?" A quantum physics and art exhibition. *Optical Engineering*, *61*(8), 081807. https://doi.org/10.1117/1.OE.61.8.081807

Degen, C. L., Reinhard, F., & Cappellaro, P. (2017). Quantum sensing. *Reviews of Modern Physics*, *89*(3), 035002. https://doi.org/10.1103/RevModPhys.89.035002

Devitt, S. J. (2016). Performing quantum computing experiments in the cloud. *Physical Review A*, *94*(3), 032329. https://doi.org/10.1103/PhysRevA.94.032329

Domnitch, E., & Gelfand, D. (2016). *Ion Hole* [Installation]. http://www.archive.olats.org/feat/04-ArtScienceCollaboration_Gelfand-Domnitch.pdf

Dowling, J. P., & Milburn, G. J. (2003). Quantum Technology: The Second Quantum Revolution. *Philosophical Transactions: Mathematical, Physical and Engineering Sciences*, *361*(1809), 1655–1674. https://www.jstor.org/stable/3559215

EPO & OECD. (2025). *Mapping the global quantum ecosystem*. European Patent Office (EPO) & Organisation for Economic Co-operation and Development (OECD). https://doi.org/10.65216/20251217-0001

Feynman, R. P. (1982). Simulating physics with computers. *International Journal of Theoretical Physics*, *21*(6–7), 467–488. https://doi.org/10.1007/BF02650179

Garcia, N., Renoust, B., & Nakashima, Y. (2020). ContextNet: Representation and exploration for painting classification and retrieval in context. *International Journal of Multimedia Information Retrieval*, *9*(1), 17–30. https://doi.org/10.1007/s13735-019-00189-4

Goethe-Institut. (n.d.). Studio Quantum. *Goethe-Institut*. Retrieved 8 October 2026, from https://www.goethe.de/prj/lqs/en/index.html?wt_sc=quantum

Gómez, I. S. (2019). Quantum entanglement: A new source of artistic creation. *Arbor*, *195*(794). Scopus. https://doi.org/10.3989/arbor.2019.794n4008

Gormley, A. (2000). *Quantum Cloud* [Galvanised steel]. https://www.antonygormley.com/works/sculpture/series/quantum-cloud/quantum-cloud

Grunwald, A. (2020). The objects of technology assessment. Hermeneutic extension of consequentialist reasoning. *Journal of Responsible Innovation*, *7*(1), 96–112. https://doi.org/10.1080/23299460.2019.1647086

Halterman, A., & Keith, K. A. (2026). Codebook LLMs: Evaluating LLMs as Measurement Tools for Political Science Concepts. *Political Analysis*, *34*(2), 188–204. https://doi.org/10.1017/pan.2025.10017

Hamido, O. C. (2022). QAC: Quantum-Computing Aided Composition. In E. R. Miranda (Ed.), *Quantum Computer Music: Foundations, Methods and Advanced Concepts* (pp. 159–195). Springer International Publishing. https://doi.org/10.1007/978-3-031-13909-3_8

Hamilton, J. (2023). A Quantum Leap Into Art. *ITNOW*, *65*(1), 54–54. https://doi.org/10.1093/combul/bwad028

Heaney, L. (2019). Quantum Computing and Complexity in Art. *Leonardo*, *52*(3), 230–235. https://doi.org/10.1162/leon_a_01572

Hsieh, H.-F., & Shannon, S. E. (2005). Three Approaches to Qualitative Content Analysis. *Qualitative Health Research*, *15*(9), 1277–1288. https://doi.org/10.1177/1049732305276687

Huffman, R. (2021, December 6). There's A Burgeoning Quantum Art Scene. *Medium*. https://medium.com/qiskit/theres-a-burgeoning-quantum-art-scene-76119cca7144

Itaboraí, P. V. (2023). *Towards Quantum Computing for Audio and Music Expression* [University of Plymouth]. https://doi.org/http://dx.doi.org/10.24382/5119

Jasanoff, S., & Kim, S.-H. (2015). *Dreamscapes of Modernity: Sociotechnical Imaginaries and the Fabrication of Power*. University of Chicago Press. https://doi.org/10.7208/chicago/9780226276663.001.0001

Javadi-Abhari, A., Treinish, M., Krsulich, K., Wood, C. J., Lishman, J., Gacon, J., Martiel, S., Nation, P. D., Bishop, L. S., Cross, A. W., Johnson, B. R., & Gambetta, J. M. (2024). *Quantum computing with Qiskit* (arXiv:2405.08810). arXiv. https://doi.org/10.48550/arXiv.2405.08810

Kirby, D. (2010). The Future is Now: Diegetic Prototypes and the Role of Popular Films in Generating Real-world Technological Development. *Social Studies of Science*, *40*(1), 41–70. https://doi.org/10.1177/0306312709338325

Kirke, A. J. (2019). Applying Quantum Hardware to non-Scientific Problems: Grover's Algorithm and Rule-based Algorithmic Music Composition. In *INTERNATIONAL JOURNAL OF UNCONVENTIONAL COMPUTING* (Vol. 14, Issues 3–4, pp. 349–374). OLD CITY

PUBLISHING INC.
Knight, P., & Walmsley, I. (2019). UK national quantum technology programme. *Quantum Science and Technology*, *4*(4), 040502. https://doi.org/10.1088/2058-9565/ab4346
Kondratyeva, O. (2025). Glass art in the age of quantum tech. *Physics World*, *38*(1), 30–33. https://doi.org/10.1088/2058-7058/38/01/19
Kroker, A., & Cook, D. (2023). The quantum revolution: Art, technology, culture. In *The Quantum Revolut.: Art, Technol., Cult.* (p. 321). University of Toronto Press. https://www.scopus.com/inward/record.uri?eid=2-s2.0-85179017269&partnerID=40&md5=5815265ee481ef1c6129209c1857163e
Kubota, A. (2025). *Quantum Computer Art Studies* (Version 1.0) [Artist's book]. https://hemokosa.com/QCA/QCAbook.pdf
Landis, J. R., & Koch, G. G. (1977). The measurement of observer agreement for categorical data. *Biometrics*, *33*(1), 159–174.
Lewis, Adam M, Cerutti, Isabella, Scudo, Petra, Travagnin, Martino, Nardo, Michela, Di Girolamo, Erica, Zaurino, Elena, Martinez Cillero, Maria, Marcantonini, Claudio, & Ascheri, Andrea. (2025). *Future directions for quantum technology in Europe: An analysis of policy questions*. Publications Office of the European Union. https://data.europa.eu/doi/10.2760/0891087
Lioret, A. (2016). Quantum Art. *Computational Aesthetics*, 5 pages. https://doi.org/10.2312/EXP.20161072
Liu, I. C. H., & Chen, H.-Y. (2026). *Roadmap to Quantum Aesthetics* (arXiv:2602.08363). arXiv. https://doi.org/10.48550/arXiv.2602.08363
Mannone, M. (2021). *Quantum Sphinx* [Paper and pencil]. https://gallery.bridgesmathart.org/exhibitions/2022-joint-mathematics-meetings/maria-mannone
Mannone, M. (2022). A musical reading of a contemporary installation and back: Mathematical investigations of patterns in Qwalala. *Journal of Mathematics and Music*, *16*(1), 80–96. Scopus. https://doi.org/10.1080/17459737.2021.1871787
Mazin, W. (2026). Quantum Fractal Art: Bringing Fractals into the Age of Quantum Computing. *Nonlinear Dynamics, Psychology, and Life Sciences*, *30*(1), 149–154.
Miranda, E. R. (2017). Guide to Unconventional Computing for Music. In *Guide to Unconv. Computing for Music* (p. 284). Springer Science+Business Media. Scopus. https://doi.org/10.1007/978-3-319-49881-2
Miranda, E. R. (Ed.). (2022). *Quantum Computing in the Arts and Humanities: An Introduction to Core Concepts, Theory and Applications*. Springer International Publishing. https://doi.org/10.1007/978-3-030-95538-0
Miranda, E. R. (2024). The advent of quantum computer music: Mapping the field. *Reports on Progress in Physics*, *87*(8), 086001. https://doi.org/10.1088/1361-6633/ad627a
Museum of Contemporary Art Tokyo. (n.d.). *Mission∞infinity | space + quantum + art*. Museum of Contemporary Art Tokyo. Retrieved 8 October 2026, from https://www.mot-art-museum.jp/en/exhibitions/mission-infinity/
Piispanen, L., Pfaffhauser, M., Wootton, J., Togelius, J., & Kultima, A. (2025). Defining quantum games. *EPJ Quantum Technology*, *12*(1), Article 1. https://doi.org/10.1140/epjqt/s40507-025-00308-7
Preskill, J. (2025). Beyond NISQ: The Megaquop Machine. *ACM Transactions on Quantum Computing*, *6*(3), 1–7. https://doi.org/10.1145/3723153
Riedel, M., Kovacs, M., Zoller, P., Mlynek, J., & Calarco, T. (2019). Europe's Quantum Flagship initiative. *Quantum Science and Technology*, *4*(2), 020501. https://doi.org/10.1088/2058-9565/ab042d

Rival Strategy. (2025, April 7). *Art and Tech in the Year of Quantum*. Rival Strategy. https://substack.com/inbox/post/160767450

Schmidt, A. (2025, August 29). Quantum Computers and Art: How and Why. *Medium*. https://medium.com/@adrian.schmidt.kit/quantum-computers-and-art-how-and-why-20d9c979be13

Schmidt, A. (under review). Towards Cultural Quantum Diplomacy: Artistic Quantum Futures through Hermeneutic Technology Assessment. In *Routledge Handbook of Quantum Diplomacy and Geopolitics*. Routledge.

Schmidt, A., Artaud, A., Aydinoglu, A. U., Bötticher, A., Araiza Bravo, R., Chiofalo, M. L. M., Coates, R., Ercan, I., Grinbaum, A., Haworth, E., Ten Holter, C., de Jong, E., Karstens, B., Kettemann, M. C., Knörr, A., Lee, C. A. L., Marco, F., Mehnert, W., Meyer, J. C., … Seskir, Z. C. (2026). Current and Future Directions for Responsible Quantum Technologies: A ResQT Community Perspective. *Ethics and Society*, *20*(3), 28. https://doi.org/10.1007/s11569-026-00522-2

Schmidt, A., Seskir, Z. C., Weinberger, N., & Coenen, C. (2025). *Technikfolgenabschätzung für die Quantentechnologien*. https://doi.org/10.5445/IR/1000185698

Selby, S. (2026). Working in Superposition: Quantum Art, Critical Practice, and Technological Power. In D. R. Upton, S. Selby, A. S. Park, & O. A. Bamgboyé, *The Q Word: A field guide to quantum art*. Gainsorough Press.

Shin, J. (2024). A Study on Artistic Creation Using Quantum Computing Technology: An Interdisciplinary Approach to Quantum Phenomena. *The Korean Society of Science & Art*, *42*(4), 189–205. https://doi.org/10.17548/ksaf.2024.09.30.189

Solano, E. (2022, October 11). *Quantum Art: Definition, Classification, Quantum NFTs* [LinkedIn-Post]. LinkedIn. https://www.linkedin.com/posts/enrique-solano-812631154_quantum-art-definition-classification-activity-6853254714675621888-8alA/

Thomas, P. (2018). *Quantum Art & Uncertainty*. Intellect. https://intellectdiscover.com/content/books/9781783209019

Thomas, P. (2026). *Concerning The Quantum in Art*. Routledge & CRC Press. https://www.routledge.com/Concerning-The-Quantum-in-Art/Thomas/p/book/9781041161943

Topel, S., Serniak, K., Burkhart, L., & Carle, F. (2022). Superconducting Qubits as Musical Synthesizers for Live Performance. In E. R. Miranda (Ed.), *Quantum Computer Music: Foundations, Methods and Advanced Concepts* (pp. 447–464). Springer International Publishing. https://doi.org/10.1007/978-3-031-13909-3_18

University of Bristol. (2017, June 7). *A beautiful hybrid: Quantum computer art*. University of Bristol. https://www.bristol.ac.uk/news/2017/june/computer-art.html

Voss-Andreae, J. (2011). Quantum Sculpture: Art Inspired by the Deeper Nature of Reality. *Leonardo*, *44*(1), 14–20.

Walker, K., & von Ompteda, K. (2016). Physicspace: From Quantum to Human Scale! *IEEE VIS*. https://researchonline.rca.ac.uk/id/eprint/2592

Wilson, S. (2001). *Information Arts: Intersections of Art, Science, and Technology*. The MIT Press. https://doi.org/10.7551/mitpress/3765.001.0001

Wootton, J. R., & Pfaffhauser, M. (2021, November 27). *Investigating the usefulness of Quantum Blur*. arXiv.Org. https://doi.org/10.5281/zenodo.10206639

Yalcinkaya, G. (2025, March 4). How Art Went Quantum. *ArtReview*. https://artreview.com/how-art-went-quantum-laure-prouvost-las-foundation-cern-ai-opinion-gunseli-yalcinkaya/?utm_source=chatgpt.com

Zetti, D. (2024). Public History: Quantum Computing as Imagined Infrastructure. *Technology and Culture*, *65*(4), 1323–1334. Scopus. https://doi.org/10.1353/tech.2024.a940472
Zheng, Z., & Yang, W. (2026). Creative control paradigms in quantum computer music: A systematic review and case practice. *The European Physical Journal Special Topics*. https://doi.org/10.1140/epjs/s11734-026-02590-7
Ziems, C., Held, W., Shaikh, O., Chen, J., Zhang, Z., & Yang, D. (2024). Can Large Language Models Transform Computational Social Science? *Computational Linguistics*, *50*(1), 237–291. https://doi.org/10.1162/coli_a_00502

# Appendix

## Data Availability

The data underlying this study, including the artworks included and their classifications, are available upon request from the corresponding author.

## Definition of the reference subset

To test the accuracy of the LLM-assisted classification, the output of a reference subset was compared to human reference classifications. The subset, consisting of 100 artworks and accounting for over 10% of the entire dataset, was selected using a *Quantum Random Number Generator* of LfD QRNG[9]. Two researchers classified the artworks of the subset independently in terms of scope, QT relation, and date. The remaining issues were resolved with a third author.

## Metrics to validate the AI-based classification

To assess the quality of the LLM-assisted screening, different agreement metrics were calculated for the three classification steps. In Phases I and III, the majority vote of three LLM classification runs was compared with the human-coded reference subset. In Phase II, the internal consistency among three repeated LLM runs was assessed by calculating Cohen's κ for each pair of runs and taking the arithmetic mean of the three values, denoted as ⟨κ⟩. Cases without a clear majority vote were excluded from the comparison with the human reference and assessed separately. A κ or ⟨κ⟩ value above 0.60 was used as a practical threshold to guide prompt refinement and quality control. All results, including values below this threshold, are reported in Table A1.

The following formula to calculate Cohen's Kappa was used

$$\kappa = \frac{\bar{P} - P_e}{1 - P_e}$$

with $\bar{P}$ being the mean observed agreement and $P_e$ the expected agreement (Cohen, 1960).

[9] https://www.lfdr.de/QRNG/

***Tab. A1*** *Metrics* $\kappa_n$ *of the three classification steps A-C and the respective phases I & III, including information about the used AI models and number of classified artworks* $n$*.* $\langle\kappa\rangle$ *represents the arithmetic mean of the run-to-run agreement between the three LLM runs, while single* $\kappa$ *the human reference checks. The indexes represent: "Without Social Media links"* $n_{w/o\ SM}$*, "only for included artworks"* $n_{(included)}$*, and "First Year"* $n_{FY}$*. The number* $n$ *for the full dataset depends on the number of artworks included in the respective screening step, as seen in Figure 1. The reference subset* $n_{sub} = 100$ *was reduced to only single artworks (*$n = 77$*) after the scope screening.*

| Prompt | Validation Phase | AI-Model | #artworks | Metric |
|---|---|---|---|---|
| (A) Scope | I: Prompt-development performance | ChatGPT 5.2 | $n_{A-I,\,sub} = 100$ | $\kappa_{A-I,\,sub} = 0.70$ |
| (A) Scope | III: Full-dataset internal consistency | ChatGPT 5.2 | $n_{A-II} = 907$<br>$n_{A-II,\,w/o\,SM} = 818$ | $\langle\kappa_{A-III}\rangle = 0.56$<br>$\langle\kappa_{A-III,\,w/o\,SM}\rangle = 0.63$ |
| (A) Scope | III: Final human-reference check | ChatGPT 5.2 | $n_{A-III,\,sub} = 100$<br>$n_{A-III,\,sub,\,w/o\,SM} = 92$ | $\kappa_{A-III,\,sub} = 0.64$<br>$\kappa_{A-III,\,sub,\,w/o\,SM} = 0.72$ |
| (B) QT relation | I: Prompt-development performance | ChatGPT 5.4 | $n_{B-I,\,sub} = 77$<br>$n_{B-I,\,sub,\,w/o\,SM} = 69$<br>$n_{B-I,\,sub,\,(included)} = 21$ | $\kappa_{B-I,\,sub} = 0.58$<br>$\kappa_{B-I,\,sub,\,w/o\,SM} = 0.55$<br>$\kappa_{B-I,\,sub,\,(included)} = 0.91$ |
| (B) QT relation | III: Full-dataset internal consistency | ChatGPT 5.4 | $n_{B-II} = 949$<br>$n_{B-II,\,w/o\,SM} = 869$<br>$n_{B-II,\,(included)} = 281$ | $\langle\kappa_{B-III}\rangle = 0.70$<br>$\langle\kappa_{B-III,\,w/o\,SM}\rangle = 0.70$<br>$\langle\kappa_{B-III,\,(included)}\rangle = 0.61$ |
| (B) QT relation | III: Final human-reference check | ChatGPT 5.4 | $n_{B-III,\,sub} = 77$<br>$n_{B-III,\,sub,\,w/o\,SM} = 69$<br>$n_{B-III,\,sub,\,(included)} = 21$ | $\kappa_{B-III,\,sub} = 0.46$<br>$\kappa_{B-III,\,sub,\,w/o\,SM} = 0.45$<br>$\kappa_{B-III,\,sub,\,(included)} = 0.76$ |
| (C) Date | I: Prompt-development performance | ChatGPT 5.5 | $n_{C-I,\,sub,\,FY} = 77$<br>$n_{C-I,\,sub,\,FY,\,w/o\,SM} = 69$ | $\kappa_{C-I,\,sub,\,FY} = 0.83$<br>$\kappa_{C-I,\,sub,\,FY,\,w/o\,SM} = 0.86$ |
| (C) Date | III: Full-dataset internal consistency | ChatGPT 5.5 | $n_{C-II,\,FY} = 923$<br>$n_{C-II,\,FY,\,w/o\,SM} = 843$ | $\langle\kappa_{C-III,\,FY}\rangle = 0.91$<br>$\langle\kappa_{C-III,\,FY,\,w/o\,SM}\rangle = 0.92$ |
| (C) Date | III: Final human-reference check | ChatGPT 5.5 | $n_{C-III,\,sub,\,FY} = 77$<br>$n_{C-III,\,sub,\,FY,\,w/o\,SM} = 69$ | $\kappa_{C-III,\,sub,\,FY} = 0.80$<br>$\kappa_{C-III,\,sub,\,FY,\,w/o\,SM} = 0.83$ |

The number of classified artworks varied across screening steps due to changes in corpus composition, the exclusion of inaccessible social media links, and analyses restricted to included QT categories.

For the first step *(A) Scope* we noticed that social media links, particularly from Instagram, could not be accessed by the LLM. We therefore calculated additional metrics excluding these cases and classified the remaining artworks manually. This increased run-to-run agreement from $\kappa = 0.56$ to 0.63 and agreement with the human reference from $\kappa = 0.64$ to 0.72.

For *(B) QT relation*, the classification proved more challenging. While agreement among repeated LLM runs reached $\kappa = 0.70$, agreement with the human reference across all categories was considerably lower ($\kappa = 0.46$). Disagreements particularly concerned the excluded categories "Classical Art", "Quantum-Inspired Art", and "Quantum Measured Art". Since the distinction between these categories was not central to our research question, we additionally assessed binary inclusion and exclusion, resulting in $\kappa = 0.79$ for both run-to-run agreement and comparison with the human reference. Within the included QT categories, agreement with the human reference reached $\kappa = 0.76$ ($n = 21$). These results indicate higher agreement for identifying QT-related artworks than for differentiating all categories, with the latter requiring additional manual review.

For *(C) Date*, three possible years were classified and reduced to the First Year (FY), following the order of year of completion, first public presentation, and first online mention. Since only FY was used in the subsequent analysis, agreement was assessed for this variable. All corresponding $\kappa$ values reached 0.80 or higher.

Overall, the agreement metrics varied among the screening steps. Binary inclusion decisions and date classifications showed relatively high agreement with the human reference. However, differentiating all QT categories proved more challenging and required additional manual review. Furthermore, agreement among repeated LLM runs indicates stability in classification rather than accuracy. Since the same human-coded reference subset was used for prompt development and final validation, these results may overestimate the performance of classifying previously unseen artworks.

These results in LLM-based classification were confirmed by hand for edge cases, which occurred particularly with the QT relation. An example of an QT-inspired artwork on the edge to Simulated QT or Generated QT Art is *CLOUD*, 2015 by Libby Heaney (Heaney, 2019). While the artwork is inspired by measurement-based QC algorithms, one can read in the artwork details that Heaney did not actually run such an algorithm on real or classical hardware. Instead a classical algorithm mimics the behaviour of measurement-based QC without actually implementing such an algorithm, which classifies the piece as QT-inspired. Although those distinctions might be uncertain in a first view, the three AI-assisted runs consistently classified the QT relation correctly. Another example is the artwork Quantum Sound, 2019 by Spencer Topel (Topel et al., 2022), which falls under QT-sourced on the edge to Generated QT Art. Topel used the tunneling signals of transmon qubits as the input for musical effects, while not actively engaging with the qubits. Also this piece has been classified correctly by ChatGPT in line with the manual classification. Those and other examples show that the results of the AI-assisted artwork classifications are good enough, both in terms of quantitative metrics as well as qualitative assessments, and were therefore used for further analysis.

## Additional figures

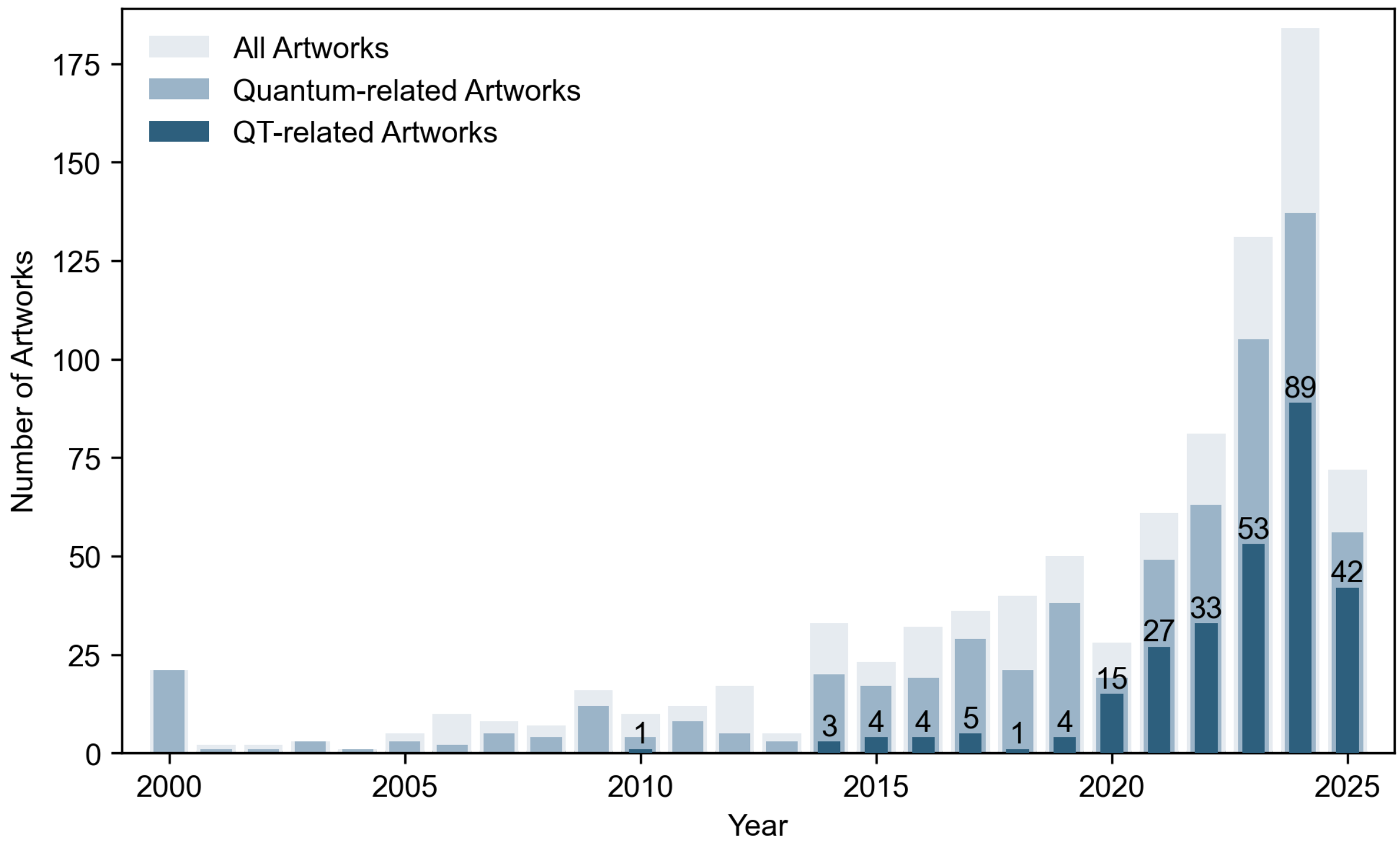

*Fig. A1* The temporal development of the entire corpus within the timeframe of the study. "Quantum-related" artworks include QT-related, as well as Quantum-Inspired and Quantum-Measured artworks. "All Artworks" also includes classical artworks.

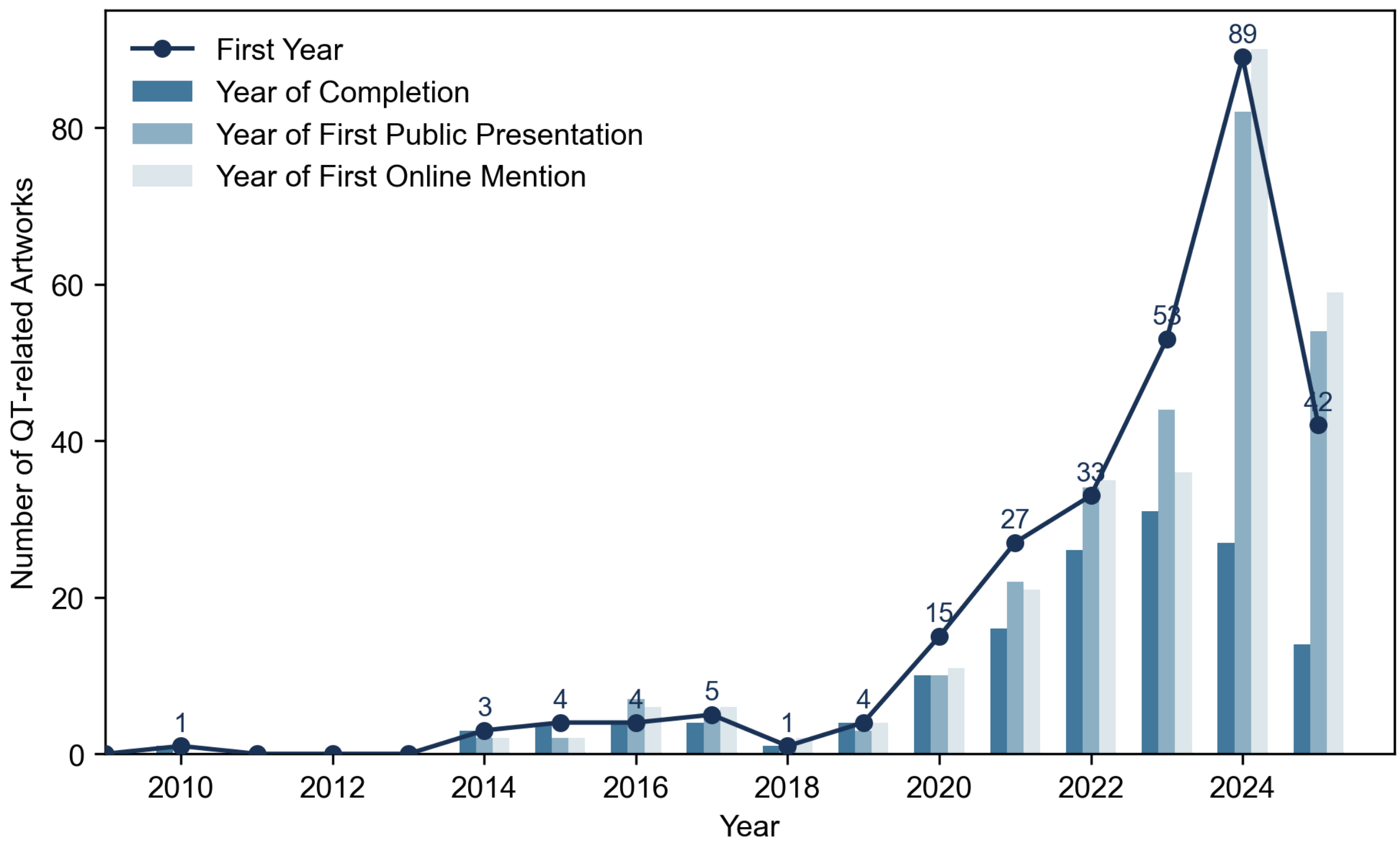

*Fig. A2* The first available year of the data classification (First Year) compared with the respective classified years. For younger dates especially, the First Year of many artworks is based on the year of *first public presentation, and sometimes on online mention, which may have shifted the First Year to a later year.*

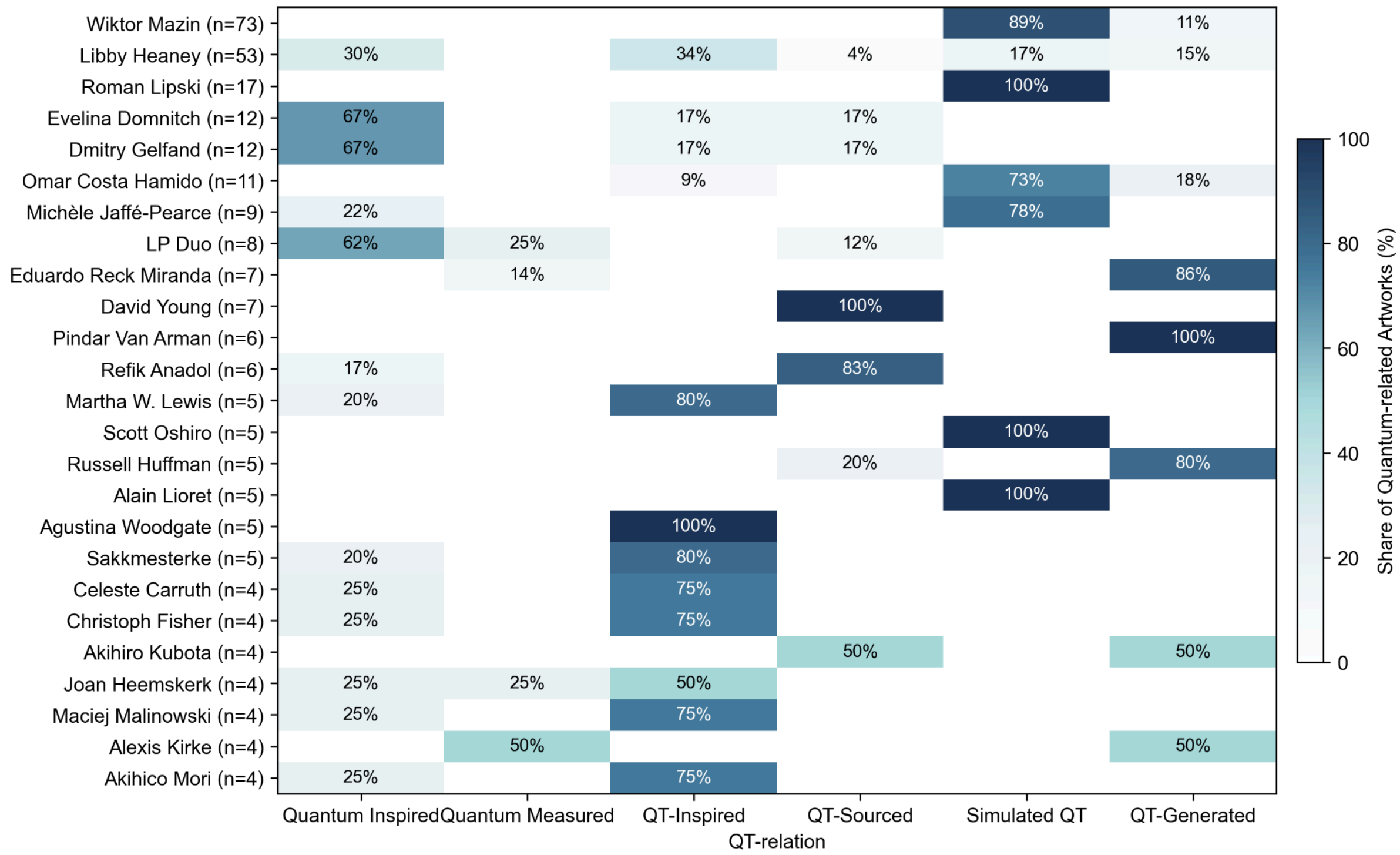

***Fig. A3*** *Share of artists' artworks for the different quantum-related categories of artists with at least 4 quantum-related artworks. n is the total number of artworks of the respective artists.*

## Tools used in QT-related artwork creation

To examine the relationships underlying the development of the active and passive use of QT in the artistic process, a separate analytical step was conducted to examine the QT tools used in these artworks. The following provides a brief description of the relevant methods and results underlying the discussion.

The 192 artworks with QT relations of "Quantum Technologies-Sourced Art," "Simulated Quantum Technologies Art," and "Quantum Technologies-Generated Art" were divided into three lists. For each QT relation, the QT tools used in the artworks were identified using ChatGPT version 5.6, the most recent available at the time of analysis. The prompt framework used for classification was adapted to output the names of the corresponding QT tools as an additional column alongside the input (number – artwork title – artist – primary link). Multiple tools could be listed per artwork. As in the screening phase, the prompts were applied to sets of ten artworks at a time. Only accessible sources were evaluated, and the results were saved.

Subsequently, the tools listed most frequently per QT relation were manually checked for correct classification. First, all tools for each artwork were separated from one another. Then, the top five were listed, and the artworks in which these tools were used were manually checked for evidence of their use. Tools with no specific connection to QT, such as Python, were excluded beforehand; only toolboxes specific to QT, such as QuTiP, were retained. For "Quantum Technologies-Sourced Art" and "Simulated Quantum Technologies Art," no distinction was made between programming languages,

Python toolboxes, or simulated and real QT hardware. For "Quantum Technologies-Generated Art," the analysis was limited to real QT hardware.

As a result, the top five tools per QT relation were identified. For "Simulated Quantum Technologies Art," the top tool is the quantum computing software stack *Qiskit* (81), followed by the quantum algorithm *Quantum Blur* (17), which can be implemented in Qiskit; *IBM Quantum* for simulated qubits (12); and *QuantumSynth*, *QAC Toolkit*, and *QuTip* (five each). For "Quantum Technologies-Sourced Art," the range extends from software stacks to actual, passively used hardware: *IBM Quantum hardware* (14), data from *Google Quantum AI Processors* (9), *Qiskit* (8), and the *Quantum Drawing Machine* (6). For "Quantum Technologies-Generated Art," *IBM Quantum hardware* clearly dominates with 26 mentions, followed by approaches using *IQM Quantum* with three mentions, the *Quandela Altair Photonic QPU* with three mentions, and the Japanese *Superconducting Quantum Computer* by *QIQB* with two mentions. It should be noted that quantum random number generators were used in two cases.